\documentclass[english,12pt]{article}
\usepackage[latin9]{inputenc}
\usepackage{mathrsfs}
\usepackage[letterpaper]{geometry}
\usepackage{amsmath,amssymb}
\usepackage{graphicx}
\usepackage{setspace}
\usepackage{babel}
\usepackage{float}

\makeatletter

\def\mathscr{\mathcal}
\def\ds{\displaystyle}

\date{}

\makeatother

\begin{document}

\title{D'Alembert sums for vibrating bar with viscous ends}
\author{Vojin Jovanovic\\
Systems, Implementation \& Integration\\
 Smith Bits, A Schlumberger Co.\\
 1310 Rankin Road\\
 Houston, TX 77032\\
e-mail: fractal97@gmail.com 
\and Sergiy Koshkin\\
Computer and Mathematical Sciences\\
University of Houston-Downtown\\
One Main Street, \#S705\\
Houston, TX 77002\\
e-mail: koshkins@uhd.edu}
\maketitle
\begin{abstract}
We describe a new method for finding analytic solutions to some initial-boundary problems for partial differential equations with constant coefficients. The method is based on expanding the denominator of the Laplace transformed Green's function of the problem into a convergent geometric series. If the denominator is a linear combination of exponents with real powers one obtains a closed form solution as a sum with finite but time dependent number of terms. We call it a d'Alembert sum. This representation is computationally most effective for small evolution times, but it remains valid even when the system of eigenmodes is incomplete and the eigenmode expansion is unavailable. Moreover, it simplifies in such cases. 

In vibratory problems d'Alembert sums represent superpositions of original and partially reflected traveling waves. They generalize the d'Alembert type formulas for the wave equation, and reduce to them when original waves can undergo only finitely many reflections in the entire course of evolution. The method is applied to vibrations of a bar with dampers at each end and at some internal point. The results are illustrated by computer simulations and comparisons to modal and FEM solutions.
\bigskip

\end{abstract}

\newpage

\section{Introduction}\label{s1}

We describe a new method for finding analytic solutions to initial-boundary problems for partial differential equations with constant coefficients, and apply the method to vibrations of a bar with two dampers at the ends and one at an internal point. We start, as in other approaches, by taking the Laplace transform with respect to time and find the Green's function of the resulting boundary eigenvalue problem. The general idea is to expand the Green's function into a series over functions with simpler dependence on the Laplace parameter, and then to invert this series termwise. In the modal approach, which is most commonly used  \cite{JK,Jovanovic}, these simpler functions are the partial fractions. Termwise inversion leads to an expansion over the eigenmodes (standing waves) of the problem, which often have explicit Laplace inverses. Our expansion also uses functions with explicit inverses, but its terms are weighted by exponents with negative real powers. The latter invert into time shifted Heaviside functions and insure that only finitely many terms of the expansion contribute to the solution at any given time. As a result, for any finite time our method produces a closed form solution as a finite sum of terms similar to those in the d'Alembert solution to the wave equation on the entire real line. We call representation of a solution as a sum of finitely many such terms, whose number may however depend on time, a {\it d'Alembert sum}. Analytically, the calculation reduces to expanding the denominator of the boundary eigenvalue problem's Green's function into a convergent geometric series. One essential restriction is that this denominator must be a linear combination of exponents with real powers. 

While this approach to Laplace inversion is common for ordinary differential equations, e.g. in signal processing 
\cite[Sec\,3.11]{SalVal}, it does not seem to have been widely used for partial differential equations. One reason is that the modal expansion is more universally applicable. Another is that one can achieve a reasonable approximation for arbitrary times with a limited number of eigenmodes, while the number of terms in a d'Alembert sum may grow rapidly with time. However, when the method works it produces an exact solution, and more importantly, it works particularly well even when the modal approach breaks down. Recall that a system of eigenmodes may be incomplete, i.e. not all functions can be approximated by linear combinations of eigenmodes, for some values of problem parameters called critical \cite{ShuBa}. In the critical cases modal expansion does not exist, but a d'Alembert sum not only exists but takes a particularly simple form.

Rather than beginning with an abstract exposition, we first show how the method works in simple examples involving a vibrating bar with dampers attached at the boundaries and at some internal point. Particular instances of this initial-boundary problem served as the main motivation for this work. Aside from being a perfect illustration, this problem provides some helpful physical insight into how and why the method works. To be more specific, let us describe the problem in more detail.
\begin{figure}[h]
\begin{centering}
\includegraphics{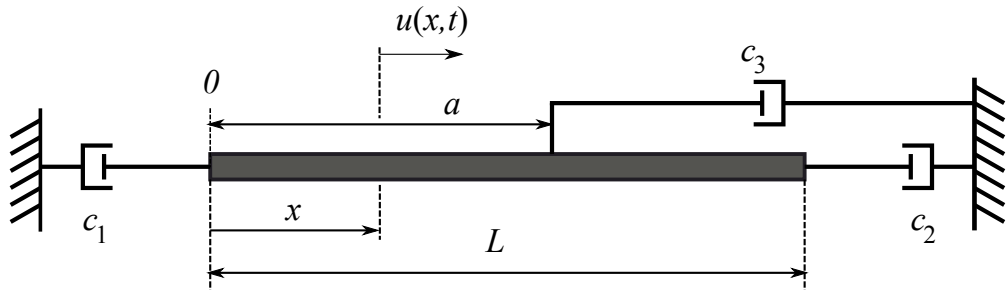} 
\par\end{centering}
\caption{\label{cap:system}A bar with viscous ends and internal damper.}
\end{figure}

Figure \ref{cap:system} depicts a bar of length $L$, free to move horizontally, suspended by two dampers
at each end and by one at the distance $a$ from its left end. Symbols
$\rho,A$ and $E$ represent the density of the bar, the constant
cross-sectional area and its modulus of elasticity respectively, the
wave speed along the bar is denoted $c:=(E/\rho)^{1/2}$. Let $c_{1}$,
$c_{2}$ and $c_{3}$ be the damping coefficients of the left, right
and internal dampers respectively, we set $h_{1}:=\frac{c}{EA}c_{1}$,
$h_{2}:=\frac{c}{EA}c_{2}$ and $h_{3}:=\frac{c}{2EA}c_{3}$ (the
extra $1/2$ simplifies some formulas). These $h_{i}$ along with
$a/L$ are the dimensionless parameters that determine qualitative
behavior of the system. Since in our case the bar can move rigidly, just
as in the problem with free ends \cite[Sec\,5.10]{Meirovitch}, we write the equations of motion
in the absolute frame that remains at rest at all times. At $t=0$
the left end of the bar is assumed to be at the origin, and $u(x,t)$
denotes the displacement of the point with initial coordinate $x$
at time $t$ in the absolute frame, see Fig.\ref{cap:system}. 

The standard derivation of the equation of motion with boundary conditions for the system without an internal damper can be found in \cite[Sec\,8.3]{Rao}. For simplicity, we do not consider spring elements attached at the boundary of the bar in this paper. The absence of springs allows the bar to move rigidly as a whole \cite[Sec\,5.10]{Meirovitch}. After considering all forces on a small element of the bar at the point of attachment of the internal damper, the equation of motion must be modified to include Dirac's delta function which models pointwise influence of the internal damper. Then, the system can be described by a modified wave equation 
\begin{equation}
u_{tt}(x,t)+2h_{3}c\,\delta(x-a)\, u_{t}(x,t)-c^{2}u_{xx}(x,t)=p(x,t),\label{eq:eom}
\end{equation}
with the boundary conditions 
\begin{equation}
u_{x}(0,t)-\frac{h_{1}}{c}\, u_{t}(0,t)=0\qquad\mbox{and}\qquad u_{x}(L,t)+\frac{h_{2}}{c}\, u_{t}(L,t)=0.\label{eq:eombc}
\end{equation}
Here $p(x,t)$ is the external force per unit mass, and the subscripts
$_x$, $_t$ denote partial derivatives with respect to space and time.
Given $u(x,t)$, the solution in the frame that moves along with the
left end of the bar is $u(x,t)-u(0,t)$. Note that $h_1$, $h_2$ and $h_3$ constants are allowed to have negative values, which correspond to control elements that supply energy rather than dampers that dissipate it. Such elements are commonly deployed as the derivative part of a proportional\mbox{-}integral\mbox{-}derivative feedback control \cite[Sec\, 3.3.2]{Ogata}. The effect of the boundary conditions on energy can be seen by differentiating the total energy $\Xi(t)=\frac{1}{2}\int_{0}^{L} \rho A u_{t}^2 dx+\frac{1}{2}\int_{0}^{L} EA u_{x}^2 dx$ of the unforced bar to obtain the energy flux
\begin{eqnarray}
\Xi_{t}(t) & = & \rho A \int_{0}^{L}u_t u_{tt} dx+\left.EA\, u_{x}u_{t}\right |_0^L -EA \int_{0}^{L} u_{t} u_{xx} dx\nonumber \\
 & = & \left.EA\, u_{x}u_{t}\right |_0^L+\int_{0}^{L} u_{t}\left[\rho A u_{tt}-EA u_{xx}\right]dx.\nonumber\\
&=&-\frac{EA}{c} \left[h_1 u_t(0,t)^2+ h_2 u_t(L,t)^2+h_3 u_t(a,t)^2\right].\label{eq:flux}
\end{eqnarray}

Eq.\eqref{eq:flux} shows that if $h_1$, $h_2$ and $h_3$ constants are positive, the energy is taken out of the system, while when they are negative the energy is pumped into it.

This problem was analyzed in \cite{JK} from the modal point of view. As long as $h_1,h_2\neq\pm1$ the problem has a countable discrete spectrum with a complete set of eigenmodes. When one or both $h_i=-1$ the solution exists only for special initial data and for finite time only, cf. \cite{UdwSuper}. In the remaining critical cases, $h_1=1$ and/or $h_2=1$, one or both boundaries become transparent, i.e. we essentially get a problem for a (semi)infinite bar with one or no boundary and no waves incoming from infinity. Since no energy is reflected back by at least one of the boundaries the standing waves can not form along the entire length of the bar, explaining why the system of eigenmodes becomes incomplete and modal expansion becomes impossible. When the standing waves can not form even on parts of the bar, closed form expressions for the time dependent Green's function were obtained in \cite{JK} by inverting the Laplace transform directly. They are combinations of Heaviside functions similar to those in the d'Alembert's solution for the entire line. However, the intermediate case, where the standing waves can form between the non-transparent boundary and the internal damper, was left untreated in \cite{JK}, because neither modal nor direct Laplace inversion worked for it. We remedy the situation in Section \ref{s3} of this paper using our new method.

Let us describe a physical picture underlying the method. Intuitively, the d'Alembert's solution and its analogs can be interpreted as tracing the evolution of an initial $\delta$-impulse as it splits into leftward and rightward traveling waves that undergo partial reflections at the dampers. The final formula represents a superposition of original and reflected traveling waves. This picture remains valid whether the standing waves can form or not, and provides a  representation complementary to the modal one. It is available even in the critical cases, including the intermediate case mentioned above. Moreover, this superposition of traveling waves reduces to d'Alembert type solutions when they exist. Of course, the number of reflections may grow indefinitely as time goes on, but at any finite time only finitely many reflected waves contribute. Thus, one obtains a closed form solution as a sum with finite but time dependent number of terms. In fact, d'Alembert type solutions exist precisely when each wave can undergo only finitely many reflections in all of time. D'Alembert sums reduce to them in such cases, hence the name. It may seem from the above description that computations eventually become intractable as the number of reflections grows with time. This would be the case under the usual method of reflections \cite[Sec\,3.13.6]{Vladimirov}, but luckily one can obtain all multiply reflected waves at once by a straightforward analytic manipulation. 

It is clear from the above discussion that d'Alambert sums will give a particularly good approximation to the solution for 'small' times, while only a small number of reflections occurred. Moreover, for the parameter values close to critical ones the reflection coefficients are small, and waves that underwent multiple reflections are strongly attenuated. This also means that one gets a good approximation by taking into account only waves reflected a small number of times. 

The paper is organized as follows. In Section \ref{s2} we apply our method to the simple case of a vibrating bar with no internal damper. Even so, the closed form solution we obtain, Eq.\eqref{eq:Grt}, appears to be new. In Section \ref{s3} we solve the intermediate case left untreated in \cite{JK}, when the internal damper is present but exactly one of the bar's boundaries is transparent. In this case the denominator of the Green's function is still a sum of two exponents. Section \ref{s4} discusses some implementation issues for computing the vibratory response using the sums of traveling waves. In particular, some care is needed when taking the time derivatives since solution formulas involve discontinuous step functions. Computer simulations and comparisons to modal and FEM solutions follow the discussion. Finally, in Section \ref{s5} we introduce general d'Alembert sums when the denominator of the Green's function is a sum of many exponents. As an illustration, the method is applied to a vibrating bar again, but with no transparent boundaries.

\section{Bar with no internal damper}\label{s2}

In this section we show how to solve initial-boundary problem Eqs.\eqref{eq:eom},\eqref{eq:eombc} by the method of d'Alembert sums in the simplest case, when no internal damper is present. Let $\mathscr{L}[\cdot]$ denote the Laplace transform over the time variable. Setting the external
force and the initial data to zero and taking the Laplace transform we get a
boundary problem for $U(x,s):=\mathscr{L}[u(x,t)]$. Since no internal damper is present $h_3=0$ in this case, and we obtain:
\begin{equation}
\frac{s^{2}}{c^{2}}\, U(x,s)-U_{xx}(x,s)=0,\label{eq:Leom}
\end{equation}
\begin{equation}
U_{x}(0,s)-h_{1}\frac{s}{c}\, U(0,s)=0\qquad\mbox{and}\qquad U_{x}(L,s)+h_{2}\frac{s}{c}\,U(L,s)=0\,.\label{eq:Leombc}
\end{equation}
We will explicitly compute the Green's function $G(x,\xi,s)$ for this problem and expand it into a series weighted by exponents with negative real powers to invert the Laplace transform.

To this end, define $\varphi(x,s),\ \psi(x,s)$ as solutions to $\frac{s^2}{c^2}\,U(x,s)-U_{xx}(x,s)=0$ satisfying only the left and the right boundary condition from Eq.\eqref{eq:Leombc} respectively. This defines them up to a constant multiple and we make them unique by normalizing $\varphi(0,s)=1=\psi(L,s)$. One easily finds that
\begin{equation}\label{eq:fi}
\varphi(x,s)=\cosh\Bigl(\frac{sx}{c}\Bigr)+h_{1}\sinh\Bigl(\frac{sx}{c}\Bigr)
=\frac12\left[(1+h_{1})e^{\frac{sx}{c}}+(1-h_{1})e^{-\frac{sx}{c}}\right]
\end{equation}
\begin{equation}\label{eq:psi}
\psi(x,s)=\cosh\Bigl(\frac{s(L-x)}{c}\Bigr)+h_{2}\sinh\Bigl(\frac{s(L-x)}{c}\Bigr)=
\frac12\left[(1+h_{2})e^{\frac{s(L-x)}{c}}+(1-h_{2})e^{-\frac{s(L-x)}{c}}\right].
\end{equation}
By definition, the Green's function $G(x,\xi,s)$ for system \eqref{eq:Leom}--\eqref{eq:Leombc} satisfies
\begin{equation}
\frac{s^2}{c^2}\,G-G_{xx}(x,s)=\delta(x-\xi),\label{eq:Gxx}
\end{equation}
\begin{equation}
G_{x}(0,\xi,s)-h_{1}\frac{s}{c}\,G(0,\xi,s)=0\qquad\mbox{and}\qquad G_{x}(L,\xi,s)+h_{2}\frac{s}{c}\,G(L,\xi,s)=0.
\label{eq:Geombc}
\end{equation}
As a function of $x$, $G$ satisfies $\frac{s^2}{c^2}\,U(x,s)-U_{xx}(x,s)=0$ for $x<\xi$ and $x>\xi$. But, by construction, any solution to $\frac{s^2}{c^2}\,U-U_{xx}=0$ satisfying the left (right) boundary condition must be a multiple of $\varphi(x,s)$ ($\,\psi(x,s)\,$). Therefore, $G(x,\xi,s)$ is equal to $A\,\varphi(x,s)$ on $[0,\xi)$ and 
$B\,\psi(x,s)$ on $(\xi,L]$ with $A$ and $B$ independent of $x$. At $x=\xi$ it is continuous, but has a jump in the first derivative to produce $\delta(x-\xi)$ in Eq.\eqref{eq:Gxx}. Namely, $G_x(\xi^+,\xi,s)-G_x(\xi^-,\xi,s)=-1$ since $G_{xx}$ enters 
Eq.\eqref{eq:Gxx} with minus. Therefore, $A,B$ can be found from the system
$$
\begin{cases}A\varphi(\xi,s)-B\psi(\xi,s)&=0\\
A\varphi'(\xi,s)-B\psi'(\xi,s)&=1\,,
\end{cases}
$$
where the primes denote the derivatives with respect to $\xi$. Solving for them in the matrix form we get
$$
\begin{pmatrix}A\\B\end{pmatrix}=\frac1{-W[\varphi,\psi]}
\begin{pmatrix}-\psi'&\psi\\-\varphi'&\varphi\end{pmatrix}\begin{pmatrix}0\\1\end{pmatrix}\,,
$$ 
where $W[\varphi,\psi]:=\left|\begin{array}{cc}\varphi & \psi\\
\varphi' & \psi'\end{array}\right|$ is the Wronskian. For convenience, we denote 
\begin{align}\label{eq:del}
\Delta(s):=-\,\frac{\,c\,}{\,s\,}\,W[\varphi,\psi]&=(1+h_{1}h_{2})\sinh\Bigl(\frac{sL}{c}\Bigr)
+(h_{1}+h_{2})\cosh\Bigl(\frac{sL}{c}\Bigr)\notag\\
&=\frac12\left[(1+h_{1})(1+h_{2})e^{\frac{sL}{c}}
-(1-h_{1})(1-h_{2})e^{-\frac{sL}{c}}\right],
\end{align}
so that explicitly $\ds{A=c\,\frac{\psi(\xi,s)}{s\Delta(s)}}$ and $\ds{B=c\,\frac{\varphi(\xi,s)}{s\Delta(s)}}$. 
We can write the Green's function in a simple form
\begin{equation}\label{eq:Glxi}
G(x,\xi,s)=\frac{c}{s\Delta(s)}\begin{cases}
\varphi(x,s)\psi(\xi,s) & x<\xi\,,\\
\psi(x,s)\varphi(\xi,s) & x>\xi\,.
\end{cases}
\end{equation}

To solve the original problem we need to invert the Laplace transform and find
$\Gamma(x,\xi,t):=\mathscr{L}^{-1}_s[G(x,\xi,s)]$. In the modal approach $G(x,\xi,s)$ is expanded into partial fractions $\frac1{(s-p)^m}$, where $p$ are the poles of $G$, then one has explicitly $\mathscr{L}^{-1}\left[\frac1{(s-p)^m}\right]=\frac{t^{m-1}}{(m-1)!}\,e^{pt}$. For $\Gamma(x,\xi,t)$ this leads to the expansion over the eigenmodes, standing waves, of the problem. Instead, we will expand $G(x,\xi,s)$ into different functions, that still can be explicitly inverted. Our expansion is obtained using the special form of the denominator of $G(x,\xi,s)$. Divide the numerator and the denominator of \eqref{eq:Glxi} by $(1+h_1)(1+h_2)e^{\frac{s}{c}L}$. Then 
\begin{equation}\label{eq:GeoSum}
G(x,\xi,s)=\frac{N(x,\xi,s)}{1-\frac{(1-h_1)(1-h_2)}{(1+h_1)(1+h_2)}\,e^{-2\frac{s}{c}L}}\,,
\end{equation}
where we combined the $x<\xi$ and $x>\xi$ cases into
\begin{equation}\label{eq:GrNum}
N(x,\xi,s)=\frac{c}{2s}\left(e^{-\frac{s}{c}|x-\xi|}+\frac{(1-h_1)}{(1+h_1)}\,e^{-\frac{s}{c}(x+\xi)}
+\frac{(1-h_2)}{(1+h_2)}\,e^{-\frac{s}{c}(2L-(x+\xi))}
+\frac{(1-h_1)(1-h_2)}{(1+h_1)(1+h_2)}\,e^{-\frac{s}{c}(2L-|x-\xi|)}\right)\!.
\end{equation}
Note that the function in Eq.\eqref{eq:GeoSum} is of the form $\ds{\frac{F(s)}{1-re^{-\alpha s}}}$. One approach to finding the inverse Laplace transform of such functions is to expand the denominator into a geometric series, convergent for sufficiently large real parts $\text{Re}(s)>0$, and to invert it termwise, see e.g. \cite[Sec\,3.11]{SalVal}. This leads to the formula
\begin{multline}\label{eq:LapGeom}
\mathscr{L}^{-1}\left[\frac{F(s)}{1-re^{-\alpha s}}\right]
=\mathscr{L}^{-1}\left[\sum_{n=0}^\infty (re^{-\alpha s})^nF(s)\right]\\
=\sum_{n=0}^\infty r^n\mathscr{L}^{-1}[e^{-\alpha ns}F(s)]
=\sum_{n=0}^\infty r^nH(t-\alpha n)\,\mathscr{L}^{-1}[F](t-\alpha n)\,.
\end{multline}
In the last equality we used the time shifting property of Laplace transforms, $\mathscr{L}^{-1}[e^{-as}F]=H(t-a)\mathscr{L}^{-1}[F](t-a)$ with $H(t)$ being the unit step Heaviside function:
\begin{equation}\label{Heaviside}
H(t):=\begin{cases}1 & t\geq0,\\0 & t<0\,.\end{cases}
\end{equation} 
In our case $F(s)=N(x,\xi,s)$, $r=\frac{(1-h_1)(1-h_2)}{(1+h_1)(1+h_2)}$ and $\alpha=\frac{2L}c$, so denoting $\Theta(x,\xi,t):=\mathscr{L}^{-1}[N(x,\xi,s)]$ we have from Eq.\eqref{eq:LapGeom}:
\begin{equation*}
\Gamma(x,\xi,t)=\sum_{n=0}^\infty\left(\frac{(1-h_1)(1-h_2)}{(1+h_1)(1+h_2)}\right)^n\!
H\Bigl(x,\xi,t-\frac{2nL}c\Bigr)\,\Theta\Bigl(x,\xi,t-\frac{2nL}c\Bigr)\,.
\end{equation*}
The $n$'th term of the sum is only non-zero for $t\geq\frac{2L}c\,n$ due to the presence of the Heaviside factors. In particular, for any fixed $t$ only finitely many terms are non-zero, and their sum gives a closed form formula for the time dependent Green's function. Let $\lfloor\cdot\rfloor$ denote the floor function which returns the largest integer not exceeding its argument. Then we can replace the infinite sum by a finite d'Alembert sum
\begin{equation}\label{eq:Grt}
\Gamma(x,\xi,t)=\sum_{n=0}^{\left\lfloor ct/2L\right\rfloor}\left(\frac{(1-h_1)(1-h_2)}{(1+h_1)(1+h_2)}\right)^n\!
H\Bigl(x,\xi,t-\frac{2nL}c\Bigr)\,\Theta\Bigl(x,\xi,t-\frac{2nL}c\Bigr)\,.
\end{equation}
Similar expressions were obtained in \cite{Veselic} by a different method. From 
Eq.\eqref{eq:GrNum} we compute
\begin{multline}\label{eq:InvNum}
\Theta(x,\xi,t)=\frac{c}{2}\left(H(ct-|x-\xi|)+\frac{1-h_1}{1+h_1}\,H(ct-(x+\xi))\right)\\
+\frac{c}{2}\frac{1-h_2}{1+h_2}\left(\,H(ct-2L+(x+\xi))
+\frac{1-h_1}{1+h_1}\,H(ct-2L+|x-\xi|)\right)\!.
\end{multline}
When the attenuation factor $\frac{(1-h_1)(1-h_2)}{(1+h_1)(1+h_2)}$ is small, only first few terms contribute significantly to the sum even for large $t$. Note that $R_i:=\frac{1-h_i}{1+h_i}$ is exactly the reflection coefficient at the $i$'th boundary, the amplitude ratio of the reflected harmonic wave to the original one 
\cite{UdwSuper}. The attenuation factor in Eq\eqref{eq:Grt} is $R_1R_2$, and it gives the amplitude decrease after a pair of reflections, one at each boundary. Note that it is $<1$ by absolute value for non-negative $h_i$. If one of the boundaries is transparent, say $h_2=1$, only the $n=0$ term in the sum survives and we get a d'Alembert type formula:
\begin{equation}\label{eq:Gmh30h21}
\Gamma(x,\xi,t)=\frac{c}{2}\left[H(ct-|x-\xi|)+\frac{1-h_{1}}{1+h_{1}}\,H(ct-(x+\xi))\right].
\end{equation}
When also $h_1=1$ and both boundaries are transparent, it reduces to the d'Alembert's solution for the entire line. Of course, even Eq.\eqref{eq:Grt} gives a closed form solution for any fixed time, but the number of terms required 
grows indefinitely as the time increases.

Thus, we have a choice between expansions into standing waves (eigenmodes) and traveling waves. In the critical cases only the latter is possible, and it behaves nicely near the critical values as well. In contrast, the modal expansion requires increasingly many terms for a good approximation near the critical values, and breaks down completely at those values. One can therefore say that the d'Alembert sum representation is complementary to the modal one from \cite{JK}.

\section{Bar with internal damper and transparent boundary}\label{s3}

It follows from the discussion in \cite{JK} that in the absence of internal damper either the Green's function has no poles at all, or the eigenmodes span the entire space. This dichotomy no longer holds when the internal damper is present. Intermediate cases arise when one of the boundaries is transparent, and it is in these cases that the d'Alembert sums become indispensable. Standing waves can still form between the damper and the non-transparent boundary, but not enough of them exist to approximate arbitrary functions. Consequently, one can find neither a modal expansion nor a d'Alembert type formula for the solution. However, a d'Alembert sum representation exists in this case as well.

As in Section \ref{s2}, it will be convenient to express the Green's function in terms of two particular solutions to the Laplace transformed equation of motion, which now becomes
\begin{equation}
\frac{s^{2}}{c^{2}}\, U(x,s)+2h_{3}\frac{s}{c}\,\delta(x-a)\, U(x,s)-U_{xx}(x,s)=0\,.\label{eq:Leomh3}
\end{equation}
Denote by $\varphi_{a}(x,s)$ ($\psi_{a}(x,s)$) the solutions
that satisfy the left (right) boundary condition only,
and are normalized to be $1$ at the corresponding boundary. Consider $\varphi_{a}$ first. 
Since the damper at $x=a$ does not affect the equation on $[0,a)$ we have $\varphi_{a}(x,s)=\varphi(x,s)$ on this
interval by definition of $\varphi(x,s)$. For $x>a$ our $\varphi_{a}$ satisfies $\frac{s^{2}}{c^{2}}\, U-U_{xx}=0$ again,
but there must be a jump in its first derivative at $a$ to accommodate
the $2h_{3}\frac{s}{c}\,\delta(x-a)\, U$ term in Eq.\eqref{eq:Leomh3}.
Along with continuity at $a$, we have $\varphi_{a}(a,s)-\varphi(a,s)=0$
and $\varphi'_{a}(a,s)-\varphi'(a,s)=2h_{3}\frac{s}{c}\varphi(a,s)$.
Since the difference $\varphi_{a}-\varphi$ also satisfies $\frac{s^{2}}{c^{2}}\, U-U_{xx}=0$
we get a Cauchy initial value problem for this function with initial point $x=a$. By inspection, 
$\varphi_{a}(x,s)-\varphi(x,s)=2h_{3}\,\varphi(a,s)\,\sinh\left(\frac{s}{c}(x-a)\right)$
for $x>a$. One can compute $\psi_{a}$ analogously or notice that by symmetry
it can be obtained from $\varphi_{a}$ by changing $x$ to $L-x$,
$a$ to $L-a$, and $h_{1}$ to $h_{2}$. With the help of the unit
step Heaviside function $H(\cdot),$ Eq.\eqref{Heaviside}, the $x<a$ and $x>a$ cases can be unified into 
\begin{align}
\varphi_{a}(x,s)=\varphi(x,s)+2h_{3}\, H(x-a)\,\varphi(a,s)\,\sinh\Bigl(\frac{s}{c}(x-a)\Bigr)\label{eq:fia}\\
\psi_{a}(x,s)=\psi(x,s)+2h_{3}\, H(a-x)\,\psi(a,s)\,\sinh\Bigl(\frac{s}{c}(a-x)\Bigr)\,.\label{eq:psia}
\end{align}
Note that $\varphi_{a}$ and $\psi_{a}$ can not be directly substituted into the second term of Eq.\eqref{eq:Leomh3} because the product $\delta(x-a)H(x-a)$ is undefined even in the sense of distributions. However, the difference 
$2h_{3}\frac{s}{c}\,\delta(x-a)\, U-U_{xx}$ does make sense because the offending product is formally canceled by the second derivative.

To further aid in notation, we define 
\begin{equation}
\Delta_{a}(s)=-\,\frac{\,c\,}{\,s\,}\,W[\varphi_{a},\psi_{a}]=\Delta(s)+2h_{3}\varphi(a,s)\psi(a,s)\,,\label{eq:trianglea}
\end{equation}
where $W[\varphi_{a},\psi_{a}]$ is the Wronskian, and is computed explicitly as 
\begin{eqnarray}
\Delta_{a}(s) & = & \left(1+h_{1}h_{2}+h_{3}(h_{1}+h_{2})\right)\sinh\Bigl(\frac{sL}{c}\Bigr)+\left(h_{1}+h_{2}+h_{3}(1+h_{1}h_{2})\right)\cosh\Bigl(\frac{sL}{c}\Bigr)\notag\label{eq:dela}\\
 &  & +h_{3}(h_{2}-h_{1})\sinh\Bigl(\frac{s(L-a)}{c}\Bigr)+h_{3}(1-h_{1}h_{2})\cosh\Bigl(\frac{s(L-a)}{c}\Bigr).
\end{eqnarray}
 We will mostly use the exponential form of this expression 
\begin{eqnarray}
\Delta_{a}(s) & = & \frac{1}{2}\left[(1+h_{1})(1+h_{2})(1+h_{3})\, e^{\frac{s}{c}L}-(1-h_{1})(1-h_{2})(1-h_{3})\, e^{-\frac{s}{c}L}\right.\notag\label{eq:delae}\\
 &  & \left.+(1-h_{1})(1+h_{2})\, h_{3}\, e^{\frac{s}{c}(L-2a)}+(1+h_{1})(1-h_{2})\, h_{3}\, e^{-\frac{s}{c}(L-2a)}\right].
\end{eqnarray}

Now we are ready to find the Green's function $G(x,\xi,s)$. By definition, it satisfies
\begin{equation}
\frac{s^2}{c^2}\,G+2h_{3}\frac{s}{c}\,\delta(x-a)\,G-G_{xx}(x,s)=\delta(x-\xi),\label{eq:Geom}
\end{equation}
\begin{equation}
G_{x}(0,\xi,s)-h_{1}\frac{s}{c}\,G(0,\xi,s)=0\qquad\mbox{and}\qquad G_{x}(L,\xi,s)+h_{2}\frac{s}{c}\,G(L,\xi,s)=0\,,
\label{eq:Geombc}
\end{equation}
and has different analytic expressions depending on relative positions of $x$, $a$ and $\xi$. Consider the case
$a<\xi$ first. As a function of $x$, $G$ satisfies Eq.\eqref{eq:Leomh3}
for $x<\xi$ and $x>\xi$. Therefore, it is equal to $A\varphi_{a}(x,s)$
on $[0,\xi)$ and $B\psi(x,s)$ on $(\xi,L]$ with $A$ and $B$ independent of $x$. At $x=\xi$ it is continuous,
but has a jump in the first derivative to produce $\delta(x-\xi)$
in Eq.\eqref{eq:Geom}. Namely, $G_{x}(\xi^+,\xi,s)-G_{x}(\xi^-,\xi,s)=-1$
because $G_{xx}$ enters Eq.\eqref{eq:Geom} with minus. Therefore,
$A,B$ can be found from the system 
\[
\begin{cases}
A\varphi_{a}(\xi,s)-B\psi(\xi,s) & =0\\
A\varphi'_{a}(\xi,s)-B\psi'(\xi,s) & =1\,,
\end{cases}
\]
where the primes denote the derivatives with respect to $\xi$. Solving for $A,B$ we get in the matrix form 
\[
\begin{pmatrix}A\\
B
\end{pmatrix}=\frac{1}{-W[\varphi_{a},\psi]}\begin{pmatrix}-\psi' & \psi\\
-\varphi'_{a} & \varphi_{a}
\end{pmatrix}\begin{pmatrix}0\\
1
\end{pmatrix}\,.
\]
But for $\xi>a$ Eq.\eqref{eq:psia} implies that $\psi_{a}(\xi,s)=\psi(\xi,s)$,
so $W[\varphi_{a},\psi_{a}]=W[\varphi_{a},\psi]=-\frac{c}{s}\Delta_{a}(s)$
by Eq.\eqref{eq:trianglea}. Therefore, 
\[
A=c\,\frac{\psi(\xi,s)}{s\Delta_{a}(s)}\qquad\text{and}\qquad B=c\,\frac{\varphi_{a}(\xi,s)}{s\Delta_{a}(s)},
\]
 so that 
\begin{equation}
G_{a<\xi}(x,\xi,s)=\frac{c}{s\Delta_{a}(s)}\begin{cases}
\varphi_{a}(x,s)\psi(\xi,s) & x<\xi\,,\\
\psi(x,s)\varphi_{a}(\xi,s) & x>\xi\,.
\end{cases}\label{eq:Galxi}
\end{equation}
Analogously, for $a>\xi$ we have 
\begin{equation}
G_{a>\xi}(x,\xi,s)=\frac{c}{s\Delta_{a}(s)}\begin{cases}
\varphi(x,s)\psi_{a}(\xi,s) & x<\xi\,,\\
\psi_{a}(x,s)\varphi(\xi,s) & x>\xi\,.
\end{cases}\label{eq:Gagxi}
\end{equation}
It will be convenient to rewrite $G$ in a form that is both
more explicit, and makes its symmetry $G(x,\xi,s)=G(\xi,x,s)$ manifest.
To this end, we introduce 
$$
g_{\varphi\psi}(x,\xi,s):=\begin{cases}
\varphi(x,s)\psi(\xi,s) & x<\xi\,,\\
\varphi(\xi,s)\psi(x,s) & x>\xi
\end{cases}\!\!\!\!\!\!,
$$ 
and compute 
\begin{multline}\label{eq:fipsi}
g_{\varphi\psi}(x,\xi,s)=\frac{1}{4}\left[(1+h_{1})(1+h_{2})e^{\frac{s}{c}(L-|x-\xi|)}
+(1-h_{1})(1+h_{2})e^{\frac{s}{c}(L-(x+\xi))}\right.\\
\left.+(1+h_{1})(1-h_{2})e^{-\frac{s}{c}(L-(x+\xi))}+(1-h_{1})(1-h_{2})e^{-\frac{s}{c}(L-|x-\xi|)}\right].
\end{multline}
Analogously, let
$$
g_{s\psi}(x,\xi,s):=\begin{cases}
\sinh\Bigl(\frac{s}{c}(x-a)\Bigr)\,\psi(\xi,s) & x<\xi\,,\\
\sinh\Bigl(\frac{s}{c}(\xi-a)\Bigr)\,\psi(x,s)& x>\xi
\end{cases}\quad\text{and}\quad 
g_{s\varphi}(x,\xi,s):=\begin{cases}
\sinh\Bigl(\frac{s}{c}(\xi-a)\Bigr)\,\varphi(x,s) & x<\xi\,,\\
\sinh\Bigl(\frac{s}{c}(x-a)\Bigr)\,\varphi(\xi,s) & x>\xi\,.
\end{cases}
$$ 
Then 
\begin{multline}\label{eq:spsi}
g_{s\psi}(x,\xi,s)=\frac{1}{4}\left[(1+h_{2})e^{\frac{s}{c}(L-a-|x-\xi|)}-(1+h_{2})e^{\frac{s}{c}(L+a-(x+\xi))}\right.\\
\left.+(1-h_{2})e^{-\frac{s}{c}(L+a-(x+\xi))}-(1-h_{2})e^{-\frac{s}{c}(L-a-|x-\xi|)}\right]\,;
\end{multline}
\begin{multline}\label{eq:sfi}
g_{s\varphi}(x,\xi,s)=\frac{1}{4}\left[(1+h_{1})e^{\frac{s}{c}((x+\xi)-a)}-(1+h_{1})e^{\frac{s}{c}(a-|x-\xi|)}\right.
\left.+(1-h_{1})e^{-\frac{s}{c}(a-|x-\xi|)}-(1-h_{1})e^{-\frac{s}{c}((x+\xi)-a)}\right].
\end{multline}
Since the $g_{\varphi\psi}$ part is common to all arrangements of $x$,
$a$ and $\xi$ we have jointly
\begin{align}
G(x,\xi,s)=\frac{c}{s\Delta_{a}(s)}\Bigl[g_{\varphi\psi}(x,\xi,s)\Bigr. & +2h_{3}\, H(x-a)H(\xi-a)\,\varphi(a,s)\, 
g_{s\psi}(x,\xi,s)\notag\label{eq:eGreen}\\
 & \ \Bigl.+\,2h_{3}\, H(a-x)H(a-\xi)\,\psi(a,s)\, g_{s\varphi}(x,\xi,s)\Bigr].
\end{align}
Note that the last two terms are non-zero only when $x$ and $\xi$
are on the same side of $a$. Therefore, whenever $a$ separates $x$
and $\xi$ the Green's function reduces to the first term.

The denominator $\Delta_{a}(s)$ is not a sum of two exponential terms as in Eq.\eqref{eq:del}, and we can not apply the same approach in general (however, see Section \ref{s5}). But, if one of the boundaries is transparent, both the numerator and the denominator simplify significantly. If say $h_2=1$, the above expressions reduce to
\begin{multline}\label{eq:Dah20}
\Delta_{a}(s) = (1+h_{1})(1+h_{3})\, e^{\frac{s}{c}L}+(1-h_{1})\,h_{3}\,e^{\frac{s}{c}(L-2a)}
=(1+h_{1})(1+h_{3})\,e^{\frac{s}{c}L}\left[1+\frac{(1-h_{1})\,h_{3}}{(1+h_{1})(1+h_{3})}\,e^{-2\frac{s}{c}a}\right]\,;
\end{multline}
\begin{equation}\label{eq:fipsih20}
g_{\varphi\psi}(x,\xi,s)=\frac{e^{\frac{s}{c}L}}{2}\left[(1+h_{1})e^{-\frac{s}{c}|x-\xi|}+(1-h_{1})e^{-\frac{s}{c}(x+\xi)}\right]\,;
\end{equation}
\begin{multline}\label{eq:spsih20}
2\varphi(a,s)\,g_{s\psi}(x,\xi,s)=\left[(1+h_{1})e^{\frac{s}{c}a}+(1-h_{1})e^{-\frac{s}{c}a}\right]
\cdot\frac{e^{\frac{s}{c}L}}{4}\left[e^{-\frac{s}{c}(|x-\xi|+a)}-e^{-\frac{s}{c}(x+\xi-a)}\right]\\
=\frac{e^{\frac{s}{c}L}}{2}\left[(1+h_{1})e^{-\frac{s}{c}|x-\xi|}-(1+h_{1})e^{-\frac{s}{c}(x+\xi-2a)}
+(1-h_{1})e^{-\frac{s}{c}(|x-\xi|+2a)}-(1-h_{1})e^{-\frac{s}{c}(x+\xi)}\right]\,;
\end{multline}
\begin{multline}\label{eq:sfih20}
2\psi(a,s)\,g_{s\varphi}(x,\xi,s)=e^{\frac{s}{c}L}\,e^{-\frac{s}{c}a}\\
\cdot\frac{1}{4}\left[(1+h_{1})e^{\frac{s}{c}((x+\xi)-a)}-(1+h_{1})e^{\frac{s}{c}(a-|x-\xi|)}
+(1-h_{1})e^{-\frac{s}{c}(a-|x-\xi|)}-(1-h_{1})e^{-\frac{s}{c}((x+\xi)-a)}\right]\\
=\frac{e^{\frac{s}{c}L}}{2}\left[(1+h_{1})e^{-\frac{s}{c}(2a-(x+\xi))}-(1+h_{1})e^{-\frac{s}{c}|x-\xi|}
+(1-h_{1})e^{-\frac{s}{c}(2a-|x-\xi|)}-(1-h_{1})e^{-\frac{s}{c}(x+\xi)}\right]\,.
\end{multline}
Analogously to Eq.\eqref{eq:GeoSum} in Section \ref{s2}, we can now represent the Green's function as
\begin{equation}\label{eq:GeoaSum}
G(x,\xi,s)=\frac{N(x,\xi,s)}{1+\frac{(1-h_{1})\,h_{3}}{(1+h_{1})(1+h_{3})}\,e^{-2\frac{s}{c}a}}\,,
\end{equation}
where (see Eq.\eqref{eq:eGreen})
\begin{equation}\label{eq:GraNum}
N(x,\xi,s)=\frac{\,c\,}{\,2(1+h_{3})\,s\,}\Bigl[n_{\varphi\psi}(x,\xi,s)+h_{3}\,H(x-a)H(\xi-a)\,n_{s\psi}(x,\xi,s)
+\,h_{3}\,H(a-x)H(a-\xi)\,n_{s\varphi}(x,\xi,s)\Bigr]\,,
\end{equation}
and
\begin{align}\label{eq:GranNum}
n_{\varphi\psi}(x,\xi,s)&:=\frac{2\cdot\,g_{\varphi\psi}}{(1+h_{1})e^{\frac{s}{c}L}}
=e^{-\frac{s}{c}|x-\xi|}+\frac{1-h_{1}}{1+h_{1}}\,e^{-\frac{s}{c}(x+\xi)}\,;\notag\\
n_{s\psi}(x,\xi,s)&:=\frac{2\cdot2\,\varphi(a,s)\,g_{s\psi}}{(1+h_{1})e^{\frac{s}{c}L}}
=e^{-\frac{s}{c}|x-\xi|}-e^{-\frac{s}{c}(x+\xi-2a)}
+\frac{1-h_{1}}{1+h_{1}}\left[e^{-\frac{s}{c}(|x-\xi|+2a)}-e^{-\frac{s}{c}(x+\xi)}\right]\,;\\
n_{s\varphi}(x,\xi,s)&:=\frac{2\cdot2\,\psi(a,s)\,g_{s\varphi}}{(1+h_{1})e^{\frac{s}{c}L}}
=e^{-\frac{s}{c}(2a-(x+\xi))}-e^{-\frac{s}{c}|x-\xi|}
+\frac{1-h_{1}}{1+h_{1}}\left[e^{-\frac{s}{c}(2a-|x-\xi|)}-e^{-\frac{s}{c}(x+\xi)}\right]\,.\notag
\end{align}
One easily verifies that whenever exponents in Eq.\eqref{eq:GraNum} have positive real parts they are multiplied by zero Heaviside factors. Thus, the inverse Laplace transform $\Theta(x,\xi,t):=\mathscr{L}^{-1}_s[N(x,\xi,s)]$ is well-defined and given by
\begin{multline}\label{eq:InvNum}
\Theta(x,\xi,t)=\frac{\,c\,}{\,2(1+h_{3})\,}\Bigl[\theta_{\varphi\psi}(x,\xi,t)+h_{3}\,H(x-a)H(\xi-a)\,\theta_{s\psi}(x,\xi,t)+\,h_{3}\,H(a-x)H(a-\xi)\,\theta_{s\varphi}(x,\xi,t)\Bigr]\,,
\end{multline}
where
\begin{align}\label{eq:GranNum}
\theta_{\varphi\psi}(x,\xi,t)&=H(ct-|x-\xi|)+\frac{1-h_{1}}{1+h_{1}}\,H(ct-(x+\xi))\,;\notag\\
\theta_{s\psi}(x,\xi,t)&=H(ct-|x-\xi|)-H(ct+2a-(x+\xi))
+\frac{1-h_{1}}{1+h_{1}}\left[H(ct-2a-|x-\xi|)-H(ct-(x+\xi))\right]\,;\\
\theta_{s\varphi}(x,\xi,t)&=H(ct-2a+(x+\xi))-H(ct-|x-\xi|)
+\frac{1-h_{1}}{1+h_{1}}\left[H(ct-2a+|x-\xi|)-H(ct-(x+\xi))\right]\,.\notag
\end{align}
Expanding $G(x,\xi,s)$ into a geometric series and inverting it termwise
we obtain a d'Alembert sum as in Eq.\eqref{eq:Grt}
\begin{equation}\label{eq:Gart}
\Gamma(x,\xi,t)=\sum_{n=0}^{\left\lfloor ct/2a\right\rfloor}\left(\frac{-(1-h_1)h_3}{(1+h_1)(1+h_3)}\right)^n\!
H(ct-2na)\,\Theta\Bigl(x,\xi,t-\frac{2na}c\Bigr)\,,
\end{equation}
where $\lfloor\cdot\rfloor$ is the floor function returning the largest integer not exceeding its argument.
The structure of $\Gamma$ is identical to that of $\Gamma$ from Eq.\eqref{eq:GraNum} with $-h_3/(1+h_3)$ playing the role of the reflection coefficient at the internal damper. By solving $-h_3/(1+h_3)=(1-h_2)/(1+h_2)$ one finds that on 
$[0,a]$ effect of the internal damper is similar to the effect of a boundary one placed at $a$ with $h_2=2h_3+1$, albeit not identical to it due to the difference in $\Theta(x,\xi,t)$.

\section{Implementation issues}\label{s4}

\begin{figure}[H]
\vspace{-0.3in}
\begin{centering}
(a)\includegraphics[scale=0.5]{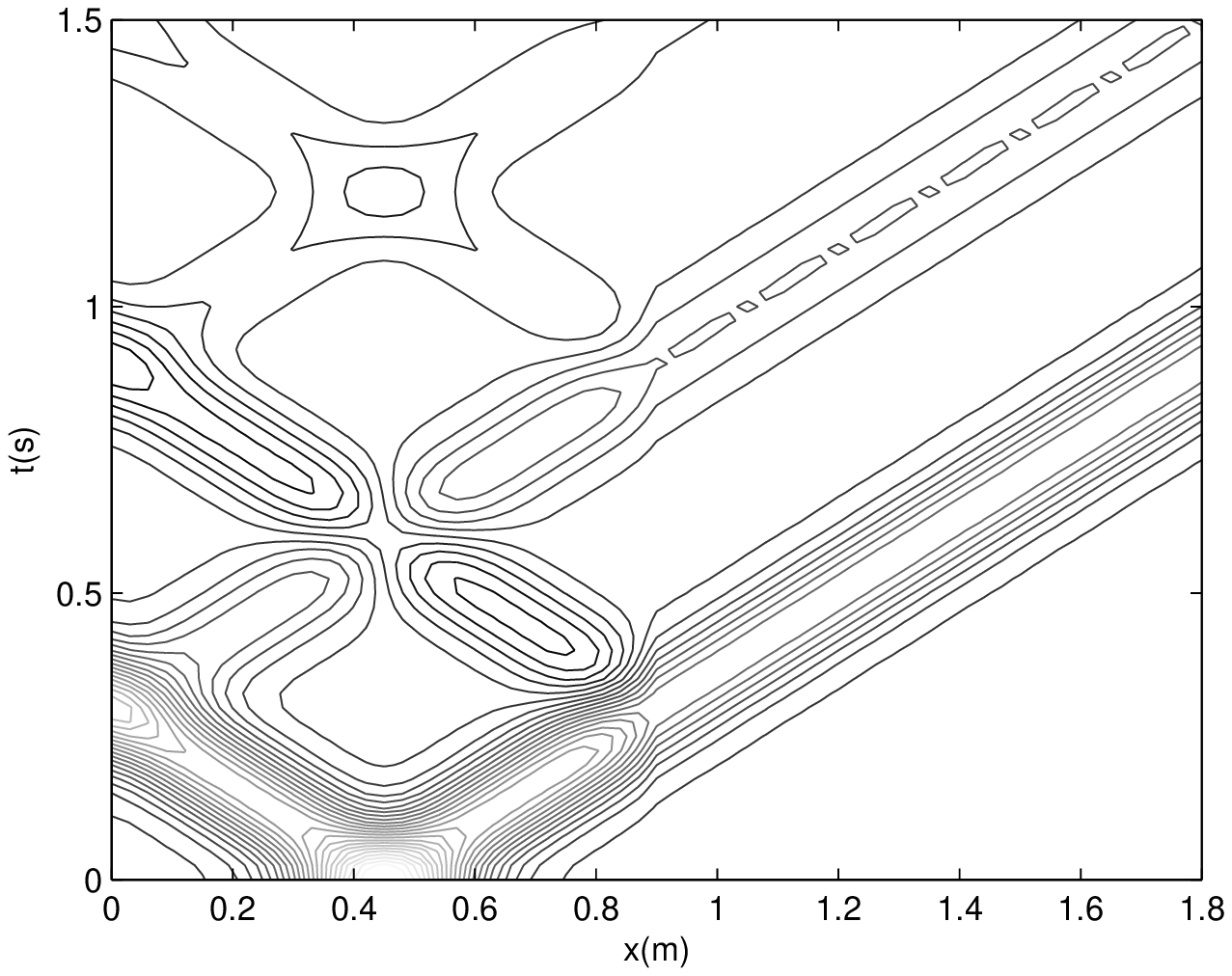}(b)\includegraphics[scale=0.5]{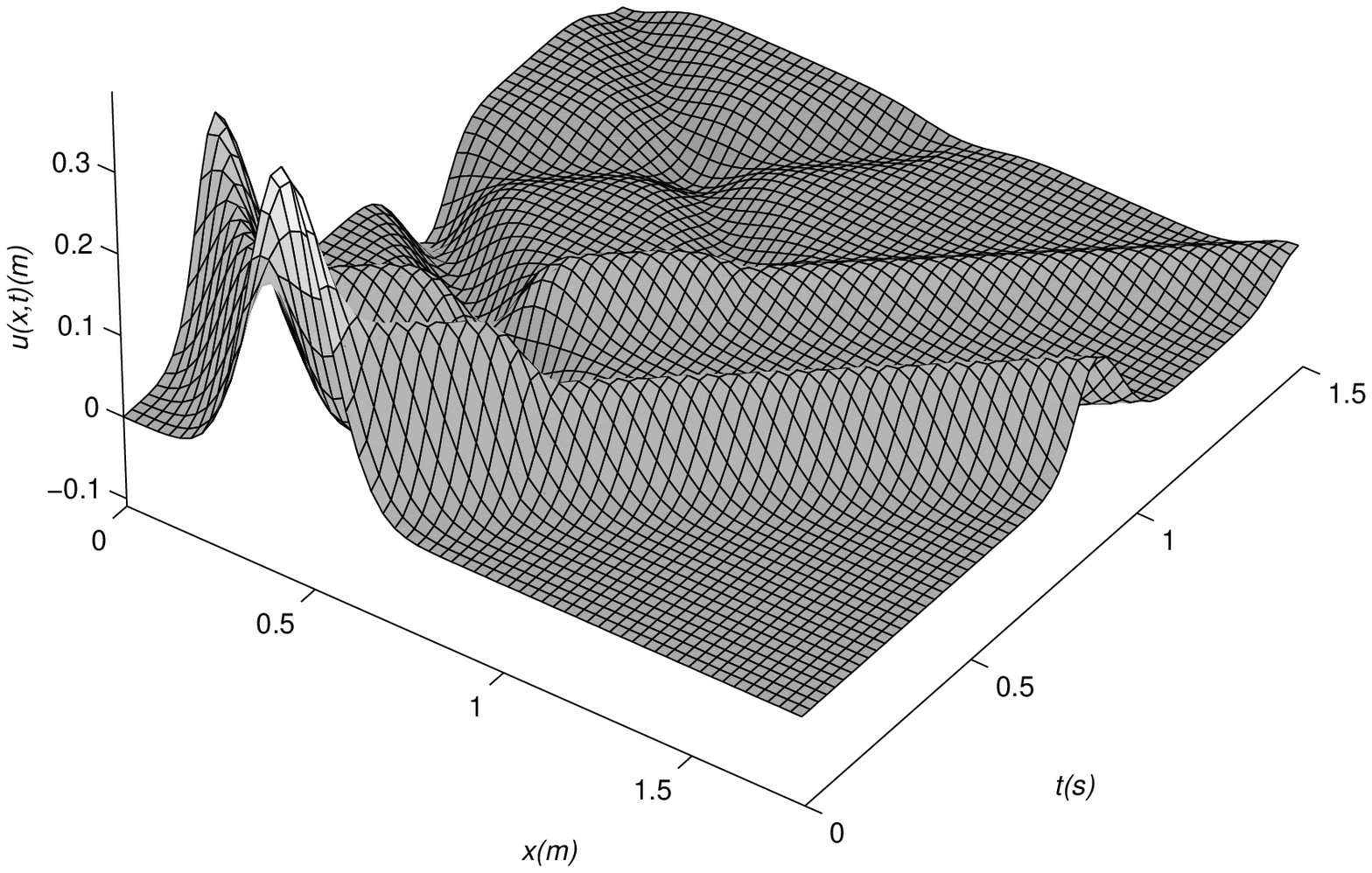} 
\par\end{centering}
\caption{\label{cap:h2eq9993D} The response of the system for $h_{1}=0.5$, $h_{2}=1$, $h_{3}=0.7$ and a Gaussian 
impulse centered at $0.25L$ where (a) is the contour plot of the response (b).}
\end{figure}

\begin{figure}[H]
\centering{}(a)\includegraphics[scale=0.5]{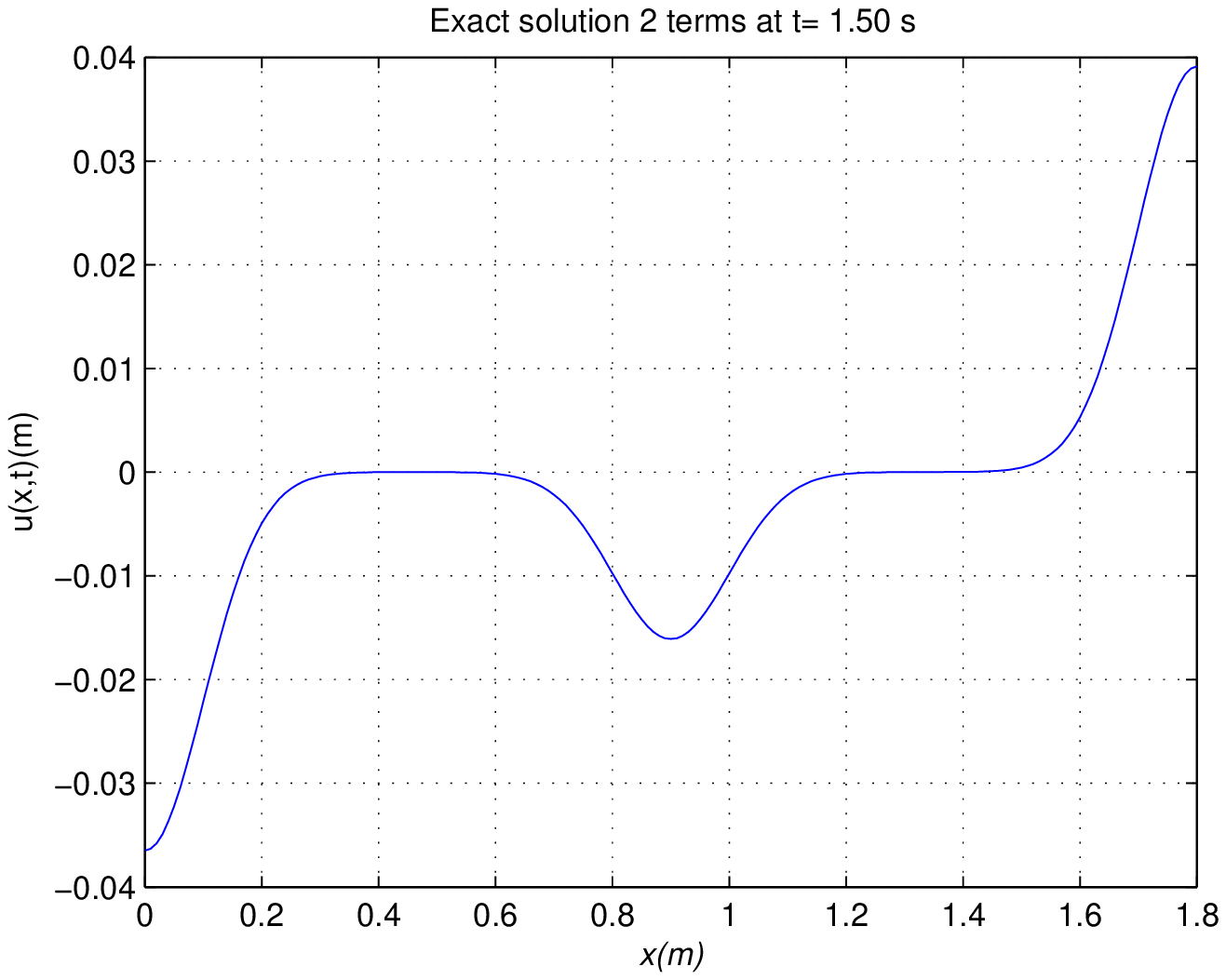}(b)\includegraphics[scale=0.5]{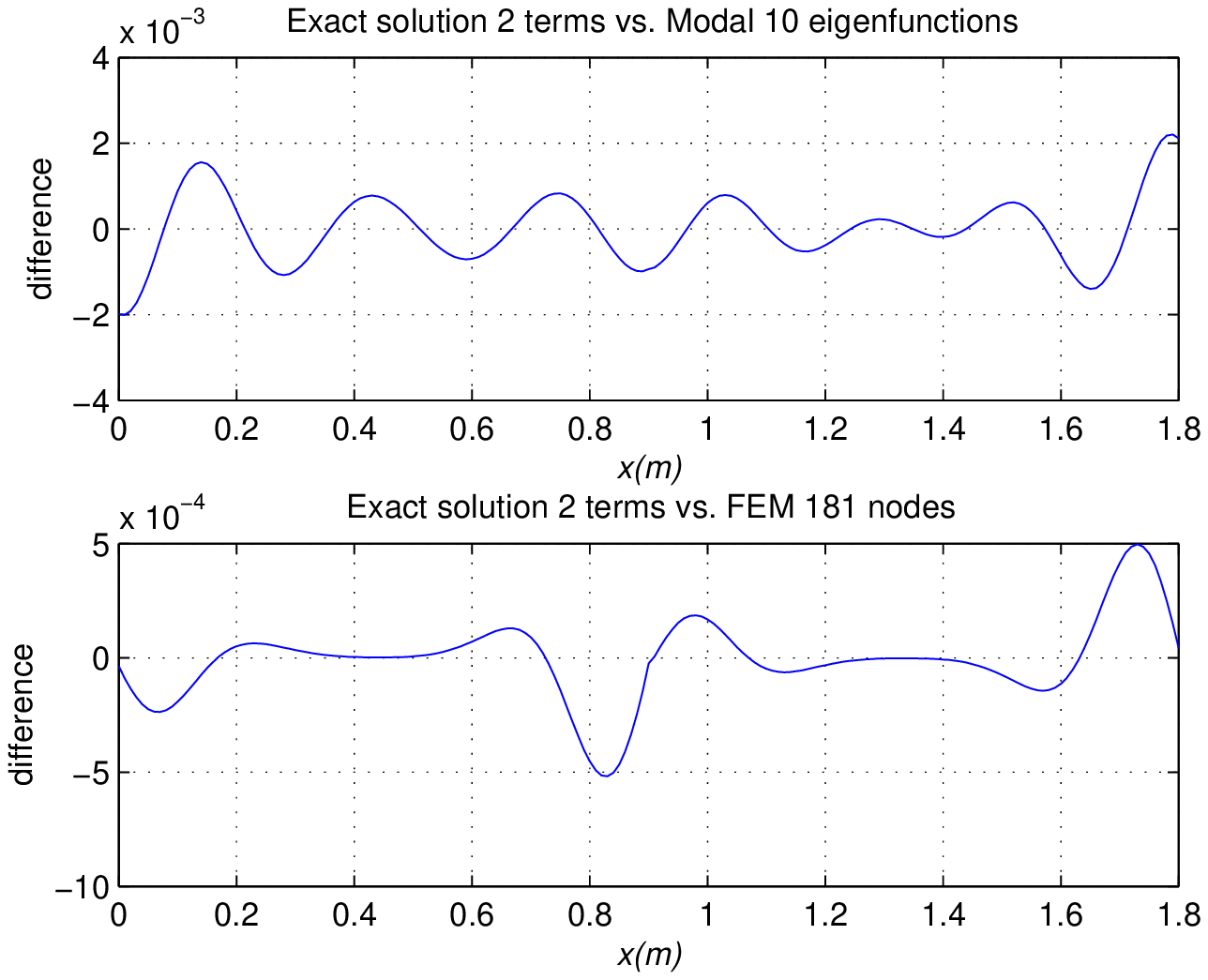}
\caption{\label{cap:h2eq999Errors} The response of the system at $t=1.5$ s with $h_{1}=0.5$ and $h_{3}=0.7$ (a),
and errors compared to modal approach (with $h_{2}=.999$) and FEM (b).}
\end{figure}

Computation of vibratory response via d'Alembert sums to presence of distributional terms in the Green's function. We address them in this section, and also compare d'Alembert sums to modal expansions and FEM solutions. The Laplace transform of the original system \eqref{eq:eom}--\eqref{eq:eombc} is 
\begin{equation}
\frac{s^{2}}{c^{2}}\, U(x,s)+2h_{3}\frac{s}{c}\,\delta(x-a)\, U(x,s)-U_{xx}(x,s)=s\,\frac{u(x,0)}{c^{2}}+\frac{u_t(x,0)+p(x,s)}{c^{2}}+2\frac{h_{3}}{c}\,\delta(x-a)\, u(x,0),\label{eq:FLeom}
\end{equation}
\begin{equation}
U_{x}(0,s)-h_{1}\frac{s}{c}\, U(0,s)=-\frac{h_{1}}{c}u(0,0)\qquad\mbox{and}\qquad U_{x}(L,s)+h_{2}\frac{s}{c}\, U(L,s)=\frac{h_{2}}{c}u(L,0).\label{eq:FLeombc}
\end{equation}
In addition to integrals with the Green's function, the solution $U(x,s)$ involves additional boundary terms due to the presence of spectral parameter in the boundary conditions, 
see \cite{JK}: 
\begin{multline}
U(x,s)=\frac{h_{1}}{c}G(x,0,s)\, u(0,0)+\frac{h_{2}}{c}G(x,L,s)\, u(L,0)+2\frac{h_{3}}{c}G(x,a,s)\, u(a,0)\\
+\int_{0}^{L}s\, G(x,\xi,s)\frac{u(\xi,0)}{c^{2}}\, d\xi+\int_{0}^{L}G(x,\xi,s)\frac{u_t(\xi,0)+p(\xi,s)}{c^{2}}
\,d\xi\,.\label{eq:Ustilde}
\end{multline}
Since $\Gamma(x,\xi,t):=\mathscr{L}^{-1}_s[G(x,\xi,s)]$ properties of Laplace transform immediately imply 
\begin{align*}
\mathscr{L}^{-1}_s[s\, G(x,\xi,s)] & =\Gamma_{t}(x,\xi,t)+\Gamma(x,\xi,0)\,\delta(t)\\
\mathscr{L}^{-1}_s[G(x,\xi,s)\, p(\xi,s)] & =\int_{0}^{t}\Gamma(x,\xi,t-\tau)\, p(\xi,\tau)\, d\tau.
\end{align*}
Inverting Eq.\eqref{eq:Ustilde}, the solution $u(x,t)$ to the initial-boundary problem Eqs.\eqref{eq:eom}-\eqref{eq:eombc}
can be written in the form 
\begin{align}
u(x,t) & =\frac{1}{c}\Bigl[h_{1}u(0,0)\,\Gamma(x,0,t)+h_{2}u(L,0)\,\Gamma(x,L,t)+2h_{3}u(a,0)\,\Gamma(x,a,t)\Bigr]\label{eq:uxt}\\
 & \quad+\frac{1}{c^{2}}\int_{0}^{L}\Bigl[\Gamma_{t}(x,\xi,t)\, u(\xi,0)+\Gamma(x,\xi,t)\,u_t(\xi,0)\Bigr]d\xi+\frac{1}{c^{2}}\int_{0}^{t}\int_{0}^{L}\Gamma(x,\xi,t-\tau)\, p(\xi,\tau)\, d\xi\, d\tau\notag
\end{align}
Since $\Gamma(x,\xi,t)$ is not a smooth function some care should be taken in computing its derivative $\Gamma_{t}$, which is distributional. If we wish $u(x,t)$ to represent the classical solution, one has to understand $\Gamma_{t}$ in Eq.\eqref{eq:uxt} as the classical part of this distributional derivative \cite[Sec\,3.12.2]{Vladimirov}. Note also that we dropped the distributional term $\Gamma(x,\xi,0)\,\delta(t)$ when writing Eq.\eqref{eq:uxt}. It reflects the jump 
at $t=0$, but is not a part of the classical solution $u(x,t)$. Furthermore, consider the traveling wave expansion Eq.\eqref{eq:Grt} for $\Gamma$. Differentiating the sum termwise and using the distributional product rule we get a sum of expressions like 
$$
H_t(ct-2nL)\,\Theta\Bigl(x,\xi,t-\frac{2nL}c\Bigr)+H(ct-2nL)\,\Theta_t\Bigl(x,\xi,t-\frac{2nL}c\Bigr).
$$
The first term gives a sequence of delta functions. As with $\Gamma(x,\xi,0)\,\delta(t)$, their contribution is purely distributional and is not a part of $u(x,t)$. Therefore, for the purposes of computing the classical solution we have 
\begin{equation}\label{eq:Grtt}
\Gamma_t(x,\xi,t)=\sum_{n=0}^\infty\left(\frac{(1-h_1)(1-h_2)}{(1+h_1)(1+h_2)}\right)^n\!
H(ct-2nL)\,\Theta_t\Bigl(x,\xi,t-\frac{2nL}c\Bigr)\,.
\end{equation}
Equation \eqref{eq:InvNum} for $\Theta(x,\xi,t)$ itself involves Heaviside functions like 
$H(ct-|x-\xi|)$, that need to be differentiated. Almost by definition $H_t(t)=\delta(t)$, where $\delta$ is the Dirac's function, but one has to be careful with formal differentiation here. For instance, $H(ct-|x-\xi|)=H(t-\frac1c|x-\xi|)$, but formally applying the chain rule on the right leads to an incorrect answer. Using distribution theory one can verify that \cite[Sec\,3.11.1]{Vladimirov}:
\begin{align*}
H_t(ct-|x-\xi|)&=c\,\delta(ct-|x-\xi|)=c\,\delta(ct-x+\xi)+c\,\delta(ct+x-\xi)\,;\\
H_t(ct-2a\pm|x-\xi|)&=c\,\delta(ct-2a\pm|x-\xi|)=c\,H(\pm ct\mp2a)\,[\delta(ct-x+\xi)+\delta(ct+x-\xi)]\,.
\end{align*}
Integration of the corresponding terms in Eq.\eqref{eq:uxt} over $\xi$ then reduces to evaluating the initial 
displacement $u(\xi)=u(\xi,0)$ at $\xi=x-ct,\,x+ct$, etc. For convenience of the reader, we give explicit formulas for convolutions of $u(\xi)$ with time derivatives of functions from Eq.\eqref{eq:GranNum}\,:
\begin{multline*}\label{eq:DerConv}
\int_{-\infty}^{\infty}\!\!\frac{\partial\theta_{\varphi\psi}(x,\xi,t)}{\partial t}\,u(\xi)\,d\xi
=c\,u(x-ct)+c\,u(x+ct)+c\,\frac{1-h_{1}}{1+h_{1}}\,u(ct-x)\,;\\
\int_{-\infty}^{\infty}\!\!\!H(\xi-a)\frac{\partial\theta_{s\psi}(x,\xi,t)}{\partial t}\,u(\xi)\,d\xi
=c\,H(x-ct-a)u(x-ct)+c\,H(x+ct-a)u(x+ct)-H(ct-x+a)u(ct-x+2a)\\
\hspace*{2in}+c\,\frac{1-h_{1}}{1+h_{1}}H(ct-2a)\,\Bigl[H(x-ct+a)u(x-ct+2a)+H(x+ct-3a)u(x+ct-2a)\Bigr]\,;\\
\int_{-\infty}^{\infty}\!\!\!H(a-\xi)\frac{\partial\theta_{s\varphi}(x,\xi,t)}{\partial t}\,u(\xi)\,d\xi
=H(ct+x-a)u(2a-(x+ct))-c\,H(ct-x+a)u(x-ct)-c\,H(a-(x+ct))u(x+ct)\\
\hspace*{2in}+c\,\frac{1-h_{1}}{1+h_{1}}H(2a-ct)\,\Bigl[H(ct-x-a)u(x-ct+2a)+H(3a-(x+ct))u(x+ct-2a)\Bigr]\,.
\end{multline*}
As an illustration of this technique, we consider the response of the system to an initial Gaussian pulse centered at $0.25L$ with $h_{1}=0.5$ and $h_{3}=0.7$.  We set $L=1.8$ m, $c=1.5$ m/s and $a=0.5L$ here and for all subsequent calculations. The right boundary is transparent, $h_{2}=1$, i.e. waves pass through it with no reflection, see Fig.\ref{cap:h2eq9993D}. The intermediate damper affects a wave traveling to the right by reducing its hight in accordance with the parameter $h_{3}$. We wish to compare the d'Alembert sums methodology to the modal and FEM approaches.
\vspace{10 mm }
\noindent Figure \ref{cap:h2eq999Errors} depicts the response of the system at $t=1.5$ s using the exact solution 
Eq.\eqref{eq:Gart} (left), and approximation errors for the modal approach \cite{JK} and FEM (right). 
The modal solution was calculated with $h_{2}=0.999$ since $h_{2}=1$ is a critical value of the system  
and the modal approach breaks down at it \cite{JK}. This underscores the generality of d'Alembert sums as they work for critical values of the system as well. Furthermore, for $t=1.5$ s one only needs to evaluate two terms in 
Eq.\eqref{eq:Gart} to obtain the exact response of the system due to the presence of the Heaviside factor.
\noindent On the other hand, for the modal approach at least ten eigenfunctions are needed to get within $0.003$ of the exact value of the response. For the FEM approach the system needs to be discretized with at least $180$ elements for the value of the response to be within $0.0005$ of the exact one. This illustrates the advantage of d'Alembert sums for small values of time. However, for larger times this advantage is lost since more and more terms would be needed in Eq.\eqref{eq:Gart}, whereas modal approach and FEM require the same computational effort.

\section{General sums of traveling waves}\label{s5}

While constructing d'Alembert sums in the previous sections we used in an essential way the fact that the denominator of the Green's function is a sum of just two exponents. In this section we will show  that the same approach applies when more than two exponents are present, albeit the sums become more cumbersome. Thus, the method can be applied to a large class of linear vibration problems. We also address analytic issues regarding termwise inversion of Laplace transforms 
in our context.

Consider a Green's function of the form $\ds{G(x,\xi,s)=\frac1s\frac{F(x,\xi,s)}{\Delta(s)}}$, where 
$\Delta(s)=\sum_{k=0}^m\widetilde{b}_ke^{\widetilde{\alpha}_ks}$ with real powers $\widetilde{\alpha}_k$. Let 
$\widetilde{\alpha}_0$ be the largest power and set $b_k:=\widetilde{b}_k/\widetilde{b}_0$, $\alpha_k:=\widetilde{\alpha}_0-\widetilde{\alpha}_k$. Dividing the numerator and the denominator of $G$ by $\widetilde{b}_0e^{\widetilde{\alpha}_0s}$ we have
\begin{equation}\label{eq:GenSum}
G(x,\xi,s)=\frac{N(x,\xi,s)}{1+\sum_{k=1}^mb_ke^{-\alpha_ks}}\,.
\end{equation}
This is exactly the form that we used in Eqs.\eqref{eq:GeoSum}, \eqref{eq:GeoaSum}, only in both of them we had $m=1$. 
Suppose further that $|\widetilde{b}_0^{-1}e^{-\widetilde{\alpha}_0s}F(x,\xi,s)|\leq C<\infty$ for $s$ with large real parts. For instance, if $F$ itself is a linear combination of exponents with coefficients and powers depending on $x$ and $\xi$, then this amounts to assuming that the real parts of their powers do not exceed $\widetilde{\alpha}_0$. 
This is always the case in initial-boundary problems regular in the sense of Tamarkin, see \cite{OrShk}. Now set 
$q(s):=\sum_{k=1}^mb_ke^{-\alpha_ks}$ and $\alpha_{\min}:=\min_k\alpha_k$, then for $s$ with large enough real parts 
$\mathrm{Re}(s)$ one has $|q(s)|\leq\max_k|b_k|e^{-\alpha_{\min}\mathrm{Re}(s)}<1$. Therefore, the denominator of Eq.\eqref{eq:GenSum} can be expanded into a geometric series that converges absolutely and uniformly for $\mathrm{Re}(s)\geq\omega$ with large 
enough $\omega>0$:
\begin{multline}\label{eq:GenSer}
G(x,\xi,s)=\sum_{n=0}^\infty(-1)^n\left(\sum_{k=1}^mb_ke^{-\alpha_ks}\right)^{\!n}\!\!N(x,\xi,s)\\
=\sum_{n=0}^\infty\,\sum_{n_1+\dots+n_m=n}\!\!\!(-1)^n\frac{n!}{n_1!\cdots n_m!}\,b_1^{n_1}\cdots b_m^{n_m}\,
e^{-(n_1\alpha_1+\dots+n_m\alpha_m)s}N(x,\xi,s)\,,
\end{multline}
where we used the multinomial formula. Recall that the inverse Laplace transform amounts to integration along a vertical line in the complex plane also with large enough real part $\omega$\,:
\begin{equation}\label{eq:InvLap}
\Gamma(x,\xi,t)=\frac1{2\pi i}\int_{\omega-i\infty}^{\omega+i\infty} G(x,\xi,s)e^{st}ds\,.
\end{equation}
Under our assumptions , $|N(x,\xi,s)|\leq\frac{C}{|s|}$ and the summations in Eq.\eqref{eq:GenSer} can be interchanged with the integral in Eq.\eqref{eq:InvLap} for $\mathrm{Re}(s)\geq\omega$. In other words, the series in 
Eq.\eqref{eq:GenSer} can be inverted termwise. Let $\Theta(x,\xi,t):=\mathscr{L}^{-1}_s[N(x,\xi,s)]$ as before, then
$$
\mathscr{L}^{-1}_s[e^{-(n_1\alpha_1+\dots+n_m\alpha_m)s}N(x,\xi,s)]
=H\Bigl(t-\sum_{k=1}^mn_k\alpha_k\Bigr)\,\Theta\Bigl(x,\xi,t-\sum_{k=1}^mn_k\alpha_k\Bigr)\,.
$$
Heaviside factors truncate the sum to finitely many terms for any given $t>0$. Indeed, we have 
$H\Bigl(t-\sum_{k=1}^mn_k\alpha_k\Bigr)=0$ for $t<\sum_{k=1}^mn_k\alpha_k$. Since 
$\sum_{k=1}^mn_k\alpha_k\geq\alpha_{\min}\sum_{k=1}^mn_k=\alpha_{\min}n$ the only non-zero terms correspond to $n\leq t/\alpha_{\min}$. Thus, a general d'Alembert sum generalizing Eqs.\eqref{eq:Grt}, \eqref{eq:Gart} is a double sum
\begin{equation}\label{eq:GenGrt}
\Gamma(x,\xi,t)=\sum_{n=0}^{\left\lfloor t/\alpha_{\min}\right\rfloor}\,\sum_{n_1+\dots+n_m=n}\!\!\!
(-1)^n\frac{n!}{n_1!\cdots n_m!}\,b_1^{n_1}\cdots b_m^{n_m}\,
H\Bigl(t-\sum_{k=1}^mn_k\alpha_k\Bigr)\,\Theta\Bigl(x,\xi,t-\sum_{k=1}^mn_k\alpha_k\Bigr)\,,
\end{equation}
where the floor function $\lfloor\cdot\rfloor$ returns the largest integer not exceeding its argument. Note that for 
Eq.\eqref{eq:GenGrt} to make sense all $\alpha_k$ must be real. This is an essential restriction on the type of problems admitting d'Alembert sum representation. 
Note also that the internal sum in Eq.\eqref{eq:GenGrt} contains potentially $m^n$ non-zero terms for every $n$, but for 
$m=1$ it reduces to a single term. This is what makes d'Alembert sums far more attractive from the computational viewpoint when the Green's function has only two exponential terms in the denominator. Still, for small $t$ evaluation of Eq.\eqref{eq:GenGrt} remains practical even when $m>1$.

In the example of Section \ref{s3} before setting $h_2=1$ we had $\widetilde{\alpha}_0=L/c$, 
$\widetilde{\alpha}_1=(L-2a)/c$, $\widetilde{\alpha}_2=-(L-2a)/c$ and $\widetilde{\alpha}_3=-L/c$. Therefore,
$m=3$ and $\alpha_1=2a/c$, $\alpha_2=2(L-a)/c$, $\alpha_3=2L/c$ and $\alpha_{\min}=\min(a,L-a)$. Hence, for the exact answer at time $t>0$ one only needs to add up terms with $n\leq\frac{ct}{2\min(a,L-a)}$. Moreover, from Eq.\eqref{eq:delae}
\begin{equation}\label{eq:ai}
b_1=\frac{(1-h_1)\,h_3}{(1+h_1)(1+h_3)},\qquad\quad
b_2=\frac{(1-h_2)\,h_3}{(1+h_2)(1+h_3)},\qquad\quad
b_3=-\frac{(1-h_1)(1-h_2)(1-h_3)}{(1+h_1)(1+h_2)(1+h_3)}\,.
\end{equation}

The function $\Theta(x,\xi,t)$ has the same structure as in Eq.\eqref{eq:InvNum} and can be obtained in the same way. Expressions for $\theta_{\varphi\psi}$, $\theta_{s\psi}$ and $\theta_{s\varphi}$ are more cumbersome in this case and we omit them here. We derived them explicitly for simulations below using a computer algebra system.

\begin{figure}[H]
\begin{centering}
(a)\includegraphics[scale=0.5]{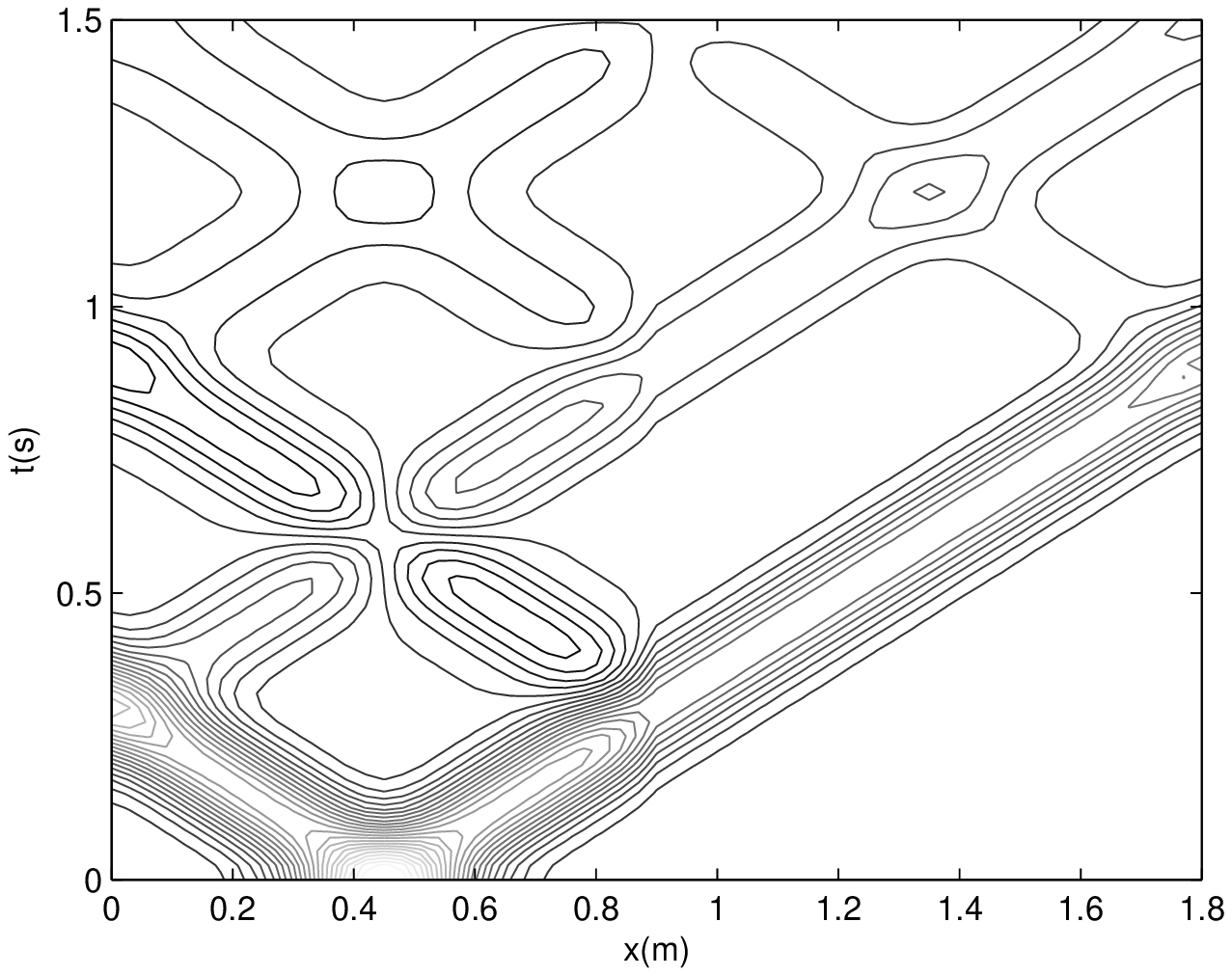}(b)\includegraphics[scale=0.5]{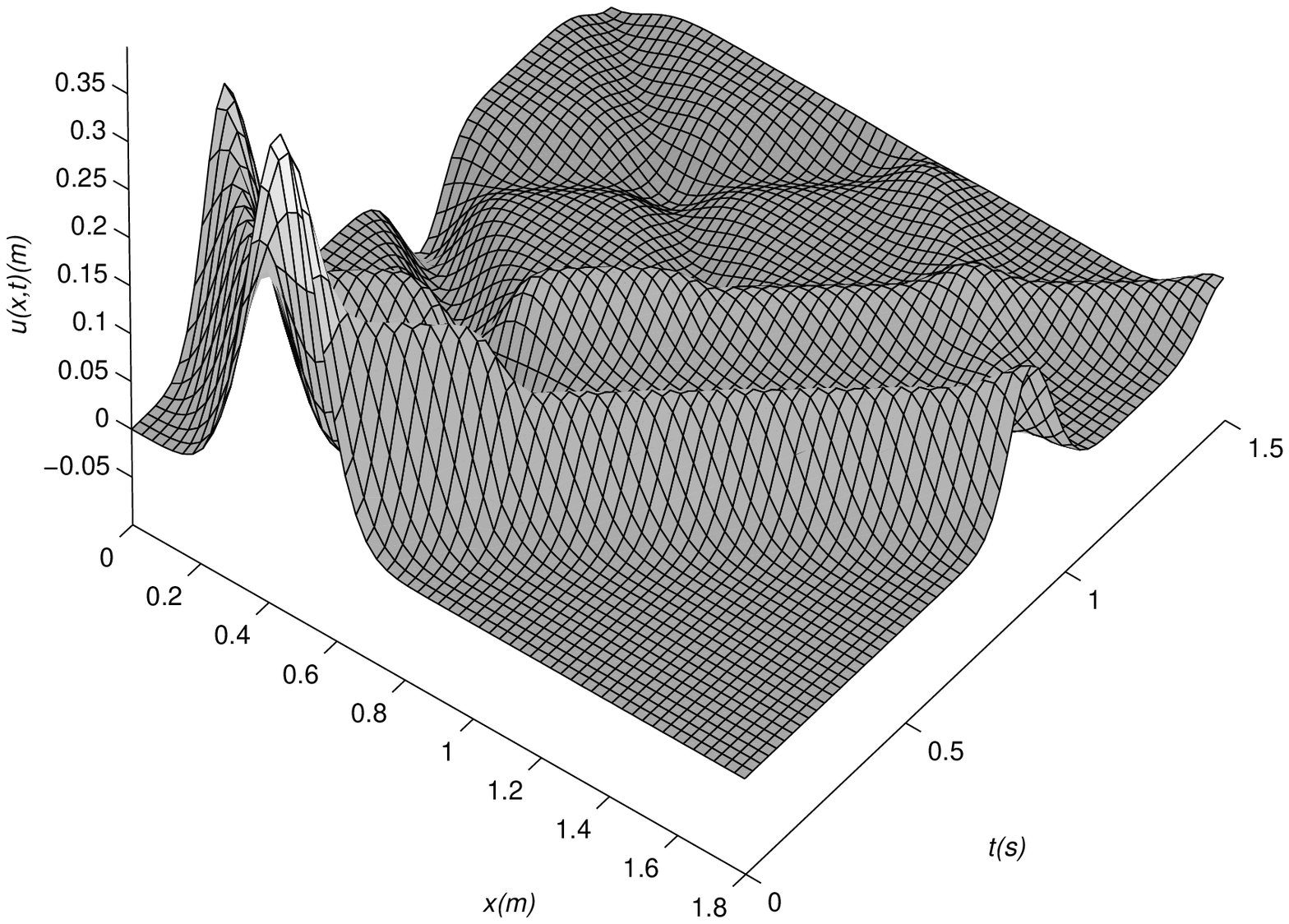} 
\par\end{centering}
\caption{\label{cap:h2eq73D}The response of the system for $h_{1}=0.5$, $h_{2}=0.7$ and  $h_{3}=0.6$, and a Gaussian impulse at $0.25L$ where (a) is the contour plot of the response (b).}
\end{figure}

\begin{figure}[H]
\centering{}(a)\includegraphics[scale=0.5]{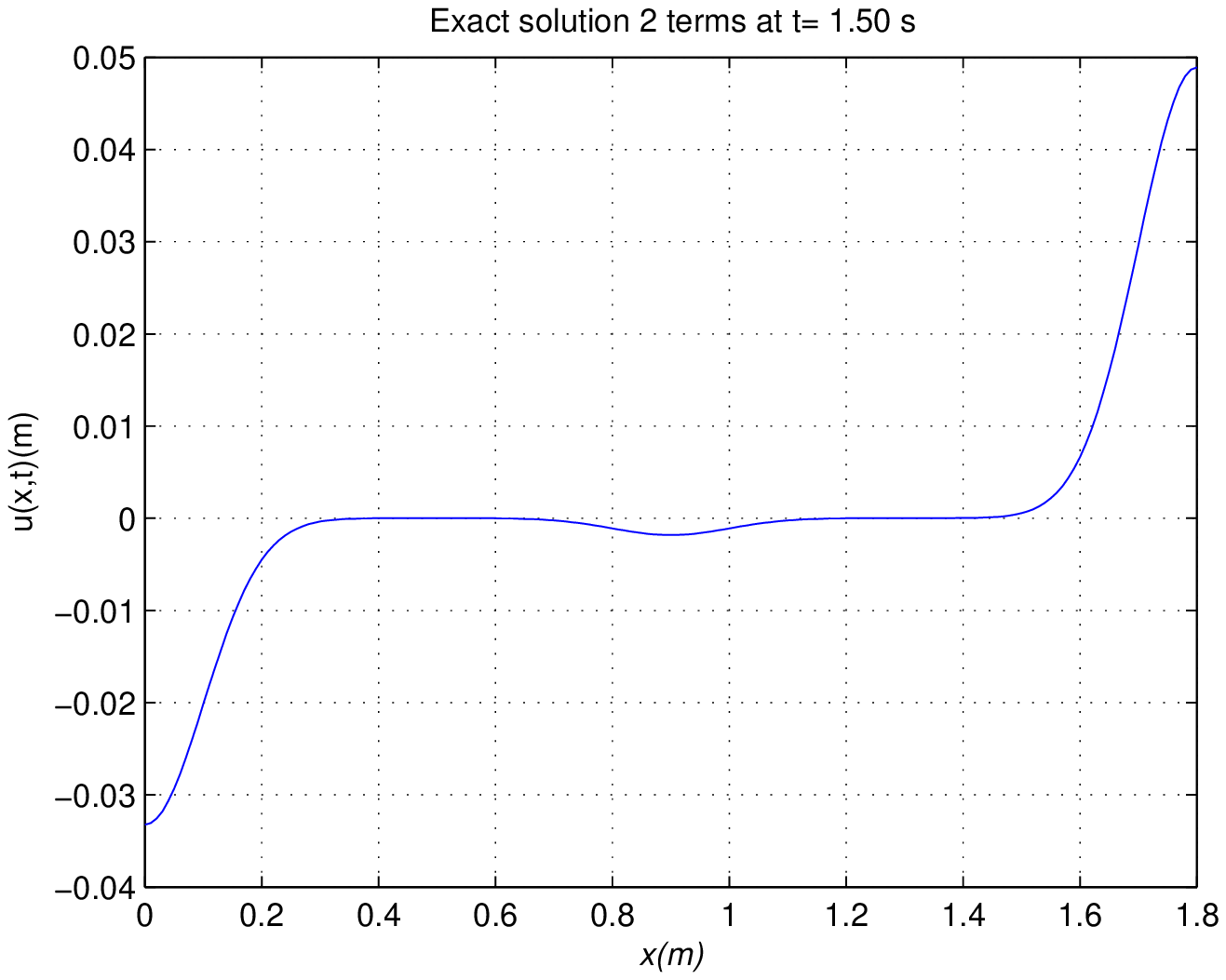}(b)\includegraphics[scale=0.5]{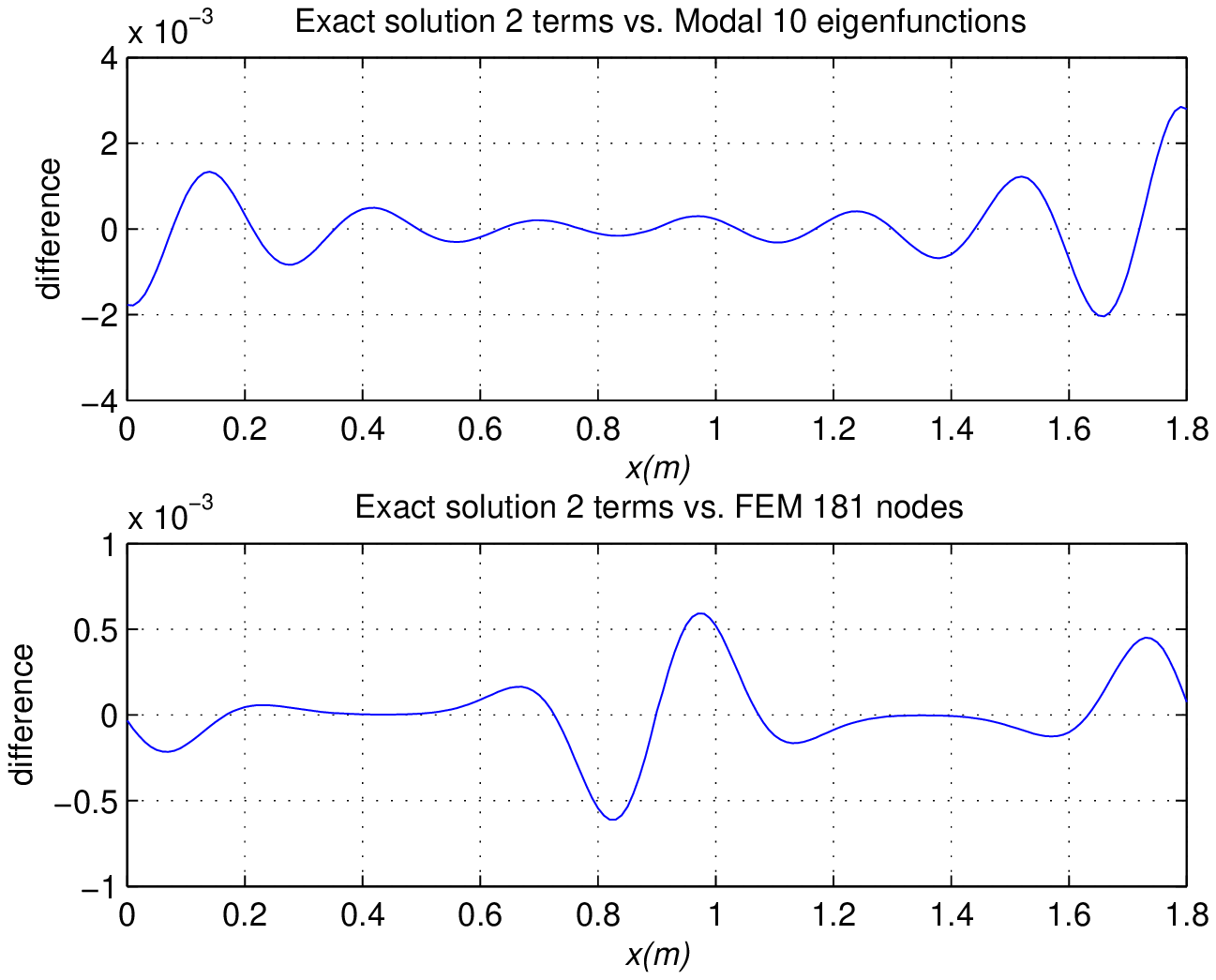}
\caption{\label{cap:h2eq7Errors} The response of the system for  $h_{1}=0.5$, $h_{2}=0.7$ and  $h_{3}=0.6$ at 
$t=1.5$ s (a), and errors compared to modal approach and FEM (b).}
\end{figure}

To illustrate the theory, consider the response of the system with $h_{1}=0.5$, $h_{2}=0.7$ and $h_{3}=0.6$ to an initial Gaussian pulse centered at $0.25L$, Fig.\ref{cap:h2eq73D}. One observes that both boundaries are not transparent 
anymore (cf. Fig.\ref{cap:h2eq9993D}). We will compare the modal and FEM solutions to d'Alembert sums. Since the value of 
$h_{2}$ is not critical here the modal approach can be applied exactly. Figure \ref{cap:h2eq7Errors} depicts the response of the system using the exact solution Eq.\eqref{eq:GenGrt} (a), and approximation errors for the modal approach \cite{JK}and FEM (b). As in Section \ref{s3}, d'Alembert sums require the least computational effort. 

\begin{figure}[H]
\begin{centering}
(a)\includegraphics[scale=0.5]{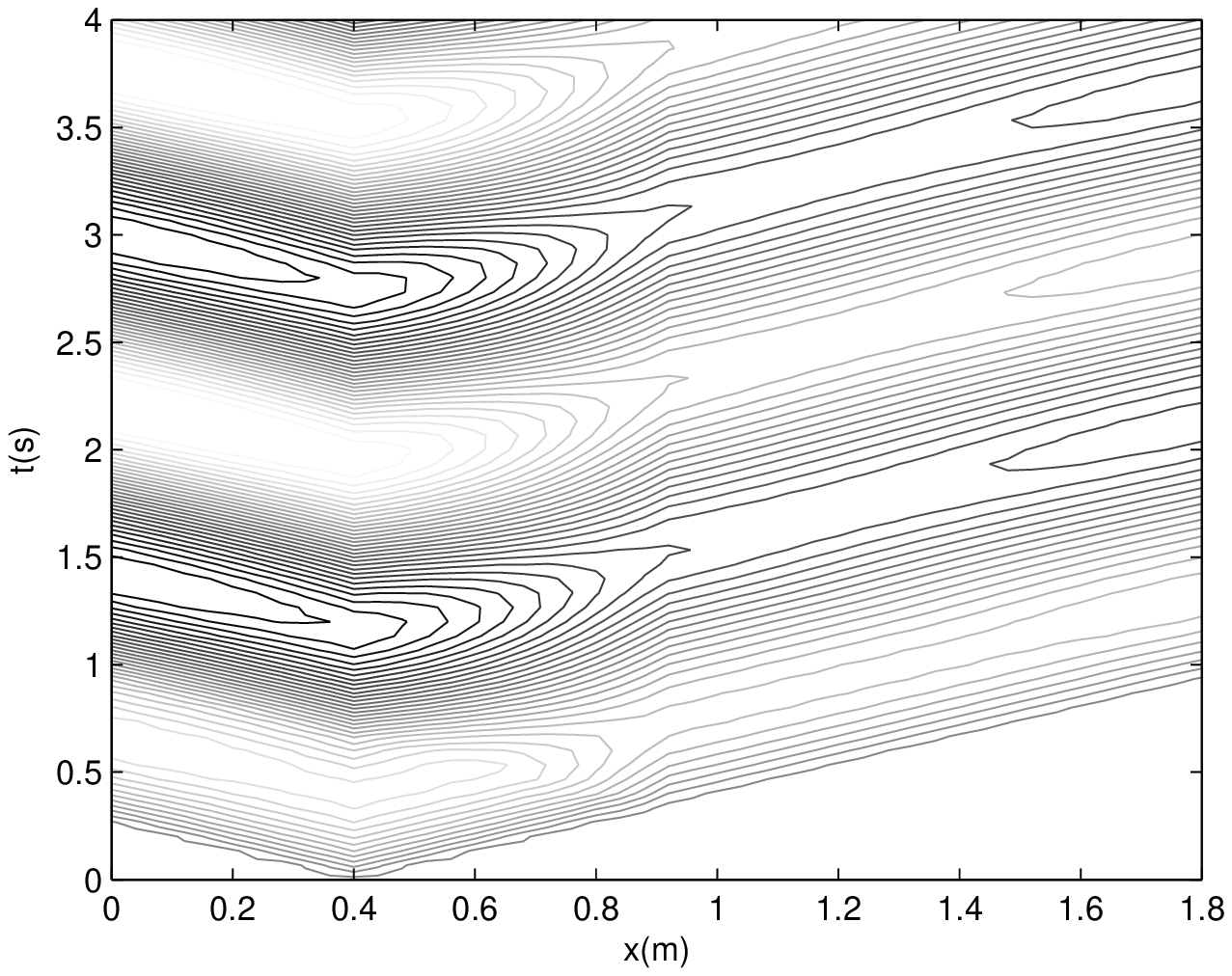}(b)\includegraphics[scale=0.5]{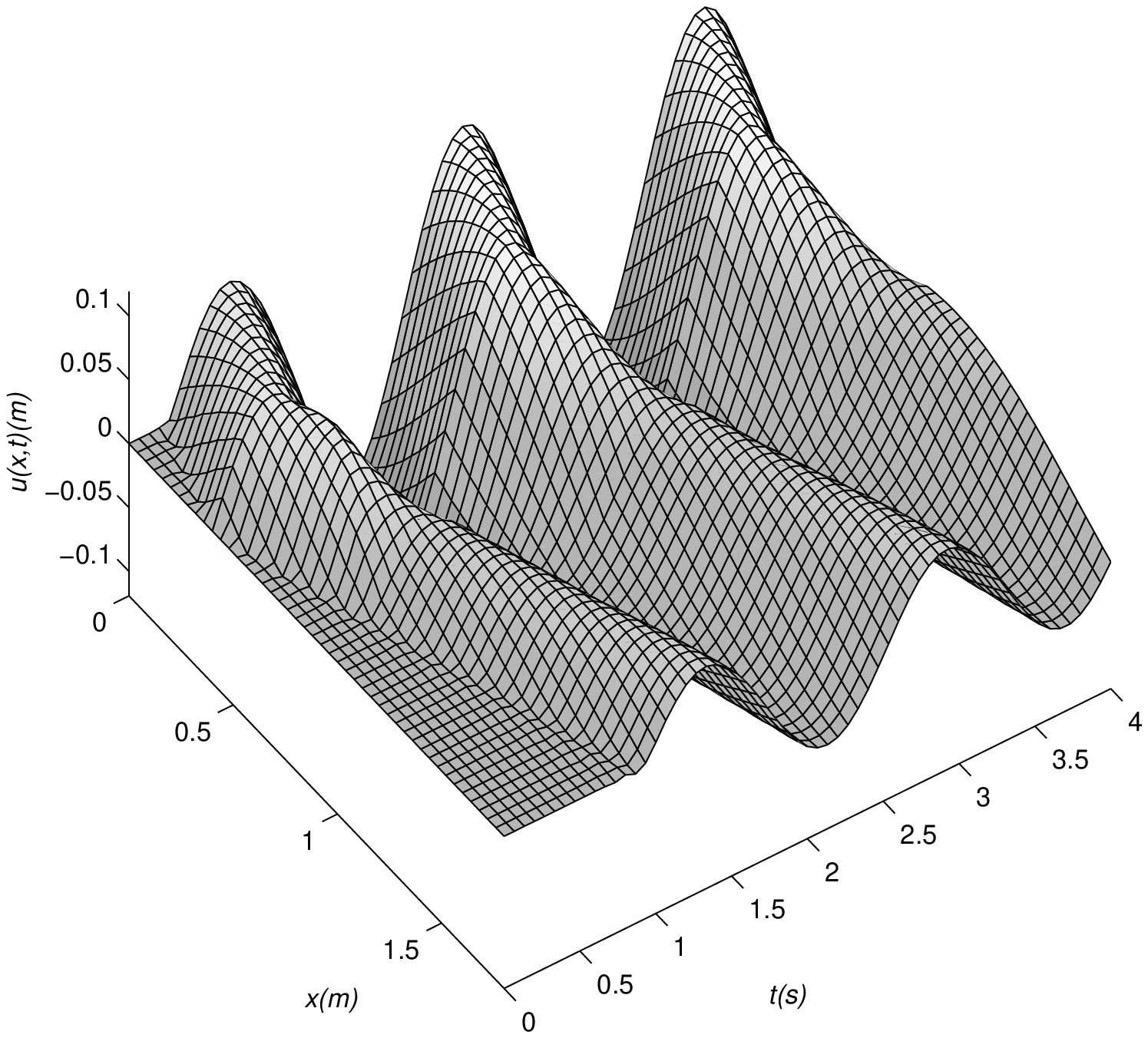} 
\par\end{centering}
\caption{\label{cap:ForceResponse3D}The response of the system subjected to a harmonic force at $0.25L$ with $h_{1}=h_{2}=0.9$ and $ h_{3}=0.6$ where (a) is the contour plot of the response (b).}
\end{figure}

\begin{figure}[H]
\begin{centering}
\includegraphics[scale=0.55]{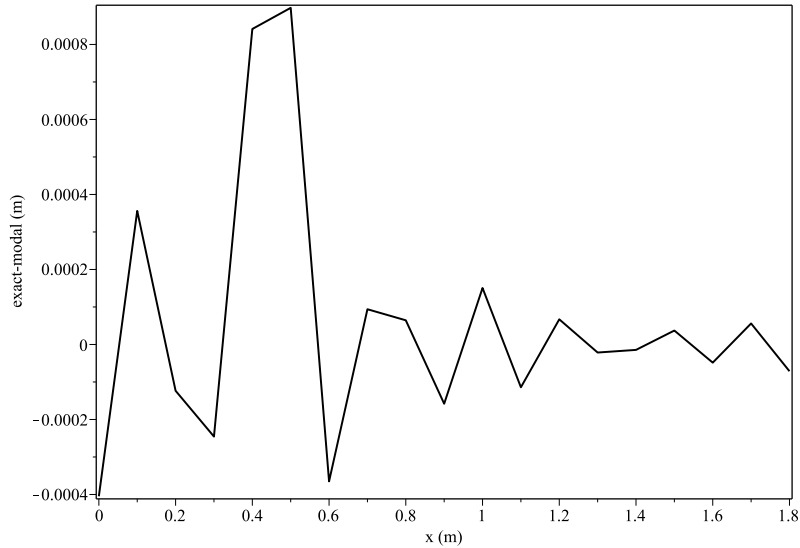} 
\par\end{centering}
\caption{\label{cap:table_ex_mod_solution}The difference between the exact and modal solution in meters.}
\end{figure}

For larger times more terms need to be retained in Eqs. \eqref{eq:Grt},\eqref{eq:Gart} and \eqref{eq:GenGrt} to get an exact solution. However, when both boundaries are almost transparent one obtains an excellent approximation by retaining just first few of their terms due to strong attenuation of reflected waves. As a result, d'Alembert sums require less computational effort than the modal or FEM approaches even for large times. This can be best observed when the system is subjected to an external force to prevent vibrations from damping out quickly. 
Figure \ref{cap:ForceResponse3D} shows the response of the system excited by an external  harmonic force at $0.25L$ with parameters $h_{1}=h_{2}=0.9$ and $h_{3}=0.6$. 
The force per unit mass is of the form $\frac{F_{0}}{\rho A}\cos(\omega t)\delta(x-0.25L)$, where $F_{0}=1$ is a constant force per unit length and $\omega=4$ is the circular frequency.
The left and right boundaries are almost transparent since the parameters $h_{1}$ and $h_{2}$ are close to 
$1$. Consequently, very little reflection occurs at the boundaries, see Fig.\ref{cap:ForceResponse3D}. 
The difference between the exact and modal solution is depicted in Figure \ref{cap:table_ex_mod_solution}. The exact solution is calculated for time $t=10$ s where for $u(x,10)$ $N=\left\lfloor t/\alpha_{\min}\right\rfloor=8$ is needed in the d'Alembert sum \eqref{eq:GenGrt}, while the modal solution $u_{e}(x,10)$ is obtained using $N_{e}=20$ eigenfunctions. One can see that to be accurate to three decimal places a large number of eigenfunctions is required. On the other hand, since the attenuation factor is small only a few first terms contribute significantly to the d'Alembert sum for all times.

\begin{figure}[H]
\begin{centering}
\includegraphics[scale=0.7]{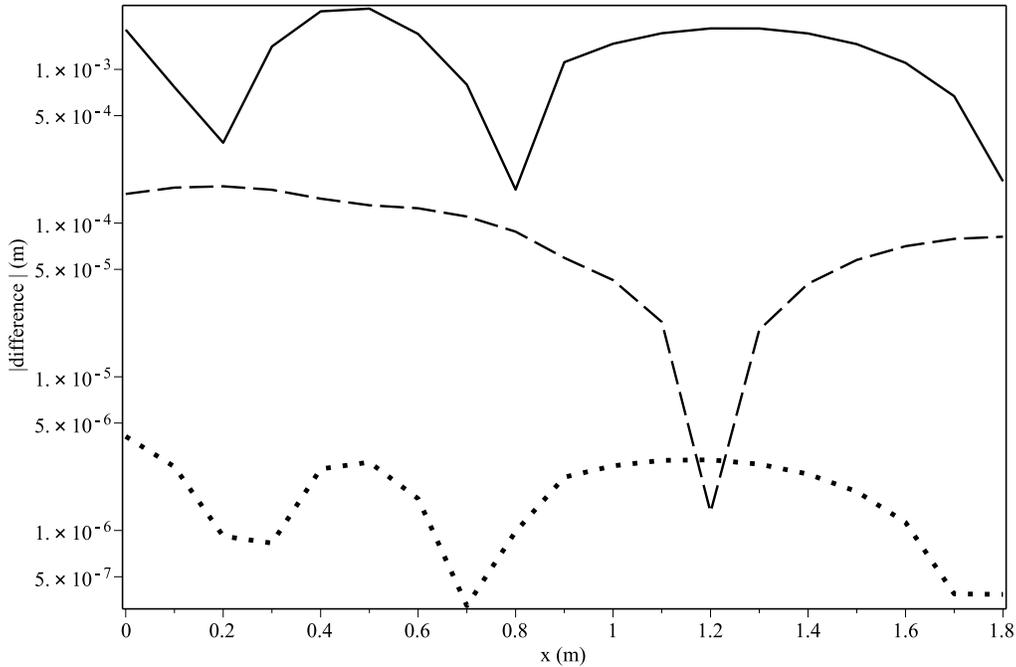} 
\par\end{centering}
\caption{\label{cap:table_errors_to_exact}Log plot of absolute difference between exact and truncated d'Alembert sums, values of are in meters.  Solid, dashed and dotted lines are represented with $N=0,1,2$ respectively. }
\end{figure}

\noindent This is depicted in Figure \ref{cap:table_errors_to_exact} where the absolute difference between the exact solution and approximations to the d'Alembert sum is truncated at first $N=0,1,2$ terms in Eq.\eqref{eq:GenGrt} is shown for time $t=10$. Already with just one term in the sum ($N=0$), the displacements obtained are within two decimal places of the exact solution. Every additional term improves the accuracy by roughly one decimal place. Therefore, for $N=1$ we already obtain the accuracy of three decimal places that required $20$ eigenfunctions under the modal approach. 

Finally, we observe that we presented numerical results in dimensional form to be consistent since this work is a continuation of \cite{JK} where the same problem with the same numerical values and units was considered.

\section{Conclusions}\label{s6}

We described the method of d'Alembert sums for solving initial-boundary problems for partial differential equations with constant coefficients, and implemented it for vibrations of a bar with boundary and internal dampers. The method presents clear computational advantages over modal expansions and FEM for small evolution times and in critical cases, when the system of eigenmodes is incomplete. In fact, since the eigenmodes are not involved at all the method is not sensitive to phenomena related to them, for instance there is no difference in its application to self-adjoint or non self-adjoint problems. However, in practice d'Alembert sums are complementary to modal expansions being ineffective when the motion is dominated by few eigenmodes and most effective when they fail to form altogether.

D'Alembert sums also hold theoretical interest providing a straightforward way to obtain closed form solutions to a number of vibratory problems. In particular, problems with multiple internal dampers and higher time derivatives in the boundary conditions can be considered. One essential limitation is that the characteristic determinant $\Delta(s)$ is a sum of real exponents, this is typically not the case for equations of order higher than two such as the biharmonic equation for beams. 

Fully analytic treatment given here will not be possible for equations with variable coefficients, but the underlying idea of inverting the numerator and the denominator of the Green's function separately is more general. Whatever exact or approximate method is used to compute the Laplace transform it might be advantageous in some problems to expand the denominator into a d'Alembert type series for inversion.

\bibliographystyle{unsrt}
\bibliography{TravWave}

\end{document}